\documentclass[preprints,article,accept,moreauthors,pdftex]{Definitions/mdpi}

\firstpage{1} 
\pubvolume{1}
\issuenum{1}
\articlenumber{0}
\pubyear{2021}
\copyrightyear{2020}
\datereceived{} 
\dateaccepted{} 
\datepublished{} 
\hreflink{} 
\pdfoutput=1

\Title{Magneto-electronic hydrogen gas sensors: a critical review}

\TitleCitation{}

\newcommand{\Ivan}{\color{black}}

\newcommand{\OKB}{\color{black}}

\usepackage{textcomp}

\Author{Ivan S. Maksymov $^{1}$ and Mikhail Kostylev $^{2}$*}

\AuthorNames{Ivan S. Maksymov and Mikhail Kostylev}

\AuthorCitation{}

\address{%
$^{1}$ \quad Optical Sciences Centre, Swinburne University of  Technology, Hawthorn, VIC 3122, Australia; imaksymov@swin.edu.au\\
$^{2}$ \quad Department of Physics and Astrophysics, University of Western Australia, Crawley, WA 6009, Australia; mikhail.kostylev@uwa.edu.au }

\corres{Correspondence: mikhail.kostylev@uwa.edu.au ; Tel.: +61-8-6488-2723}

\abstract{Devices enabling early detection of low concentrations of leaking hydrogen and precision measurements in a wide range of hydrogen concentrations in hydrogen storage systems are essential for the mass-production of fuel-cell vehicles and, more broadly, for the transition to the hydrogen economy. Whereas several competing sensor technologies are potentially suitable for this role, ultra-low fire-hazard, contactless and technically simple {\OKB magneto-electronic} sensors stand apart because they have been able to detect the presence of hydrogen gas in a range of hydrogen concentrations from 0.06\% to 100\% at atmospheric pressure with the response time approaching the industry gold standard of one second. This new kind of hydrogen sensors is the subject of this review article, where we inform the academic physics, chemistry, material science and engineering communities as well as industry researchers about the recent developments in the field of magneto-electronic hydrogen sensors, {\OKB including those based on magneto-optical Kerr effect, anomalous Hall effect and Ferromagnetic Resonance with a special focus on Ferromagnetic Resonance (FMR) based devices.} In particular, we present the physical foundations of magneto-electronic hydrogen sensors and we critically overview their advantages and disadvantages for applications in the vital areas of the safety of hydrogen-powered cars and hydrogen fuelling stations as well as hydrogen concentration meters, including those operating directly inside hydrogen-fuelled fuel cells. We believe that this review will be of interest to a broad readership, also facilitating the translation of research results into policy and practice.}


\begin{document}
\section{Introduction}
Civilisation as we know it, and, in particular, the modern society, has been built around using fossil fuels such as coal, oil and natural gas. All containing carbon and formed billions years ago as a result of geologic processes acting on the remains of organic matter, fossil fuels at present supply about 85\% of world energy. However, if we keep relying on them at the current rate, it is estimated that they will be depleted by 2060 – in less than 300\,years since the dawn of the industrial age. The use of fossil fuels also raises serious environmental and ethical concerns. In fact, the burning of fossil fuels produces around 35\,billion tonnes of carbon dioxide per year \cite{Rit20}. Therefore, the issues of global climate change and air pollution as well as their consequences such as food insecurity, political instability, terrorism and armed conflicts have pushed clean energy up the global agenda.

Nowadays, only about 15\% of world energy is supplied by non-fossil sources, including nuclear, hydroelectric and renewable energy technologies exploiting geothermal, solar, tidal and wind energies \cite{BP}. However, the world is united in the commitment to decrease the reliance on fossil fuels and increase and diversify the renewable energy supplies. Accelerating the transition to renewable energy-based economies represents a unique opportunity to stop the climate change also achieving stable economic growth, creating new employment opportunities and enhancing human welfare \cite{Byr14, Rai19}. 

Unfortunately, whereas many of non-fossil energy technologies can supply clean energy, many of them also have technological and fundamental limitations and disadvantages, including higher complexity and cost compared with traditional technologies, intermittency (e.g. solar and wind energy) and therefore the need for energy storage, major safety concerns (e.g. nuclear energy) and also geographical limitations \cite{Byr14, Rai19}. As a result, novel clean energy technologies are required and the search for them is shaping research efforts in many fields of science and technology.

{\Ivan Hydrogen is the most common chemical in the universe that can be used either in its own right or it can add value to other materials. Hence,} it has many applications such as fuel for transport or heating, a way to store and transport energy produced using a renewable energy technology or as a raw material in industrial processes \cite{Hac18}.

Hydrogen can also be the fuel in a fuel cell that produces electricity with high efficiency \cite{Hac18}, which opens up exciting opportunities for car manufacturers \cite{Pas11}. Unlike fully electric and plug-in hybrid vehicles, hydrogen cars produce the electricity themselves since they have their own efficient power plant on board--the fuel cell. A commercially-sold hydrogen-powered electric car can travel a distance of more than 500\,km on a single tank but the time required to refuel it is just five minutes compared with more than 30\,minutes needed to recharge the battery of a battery-powered electric vehicle. As a result, hydrogen-powered cars can be refuelled in dedicated service stations similarly to gasoline-powered cars. Furthermore, the weight of Li-ion batteries becomes prohibitively large for battery-powered trucks. Therefore, there is a consensus in the community that hydrogen is the only option for environmentally friendly trucks \cite{Trucks, NACFE}. Thus, given a significant greenhouse gas emission footprint from a typical passenger vehicle and a much larger one from a truck, and an estimated 1\,billion cars and 0.25\,billion trucks and buses on the road worldwide, it has been envisioned that hydrogen-based technologies could make significant contributions to decarbonisation of fossil fuel-intensive industries, thereby promoting the development of the hydrogen economy \cite{Pas11, Byr14, Rai19}.
\begin{figure}[t]
\includegraphics[width=8.5 cm]{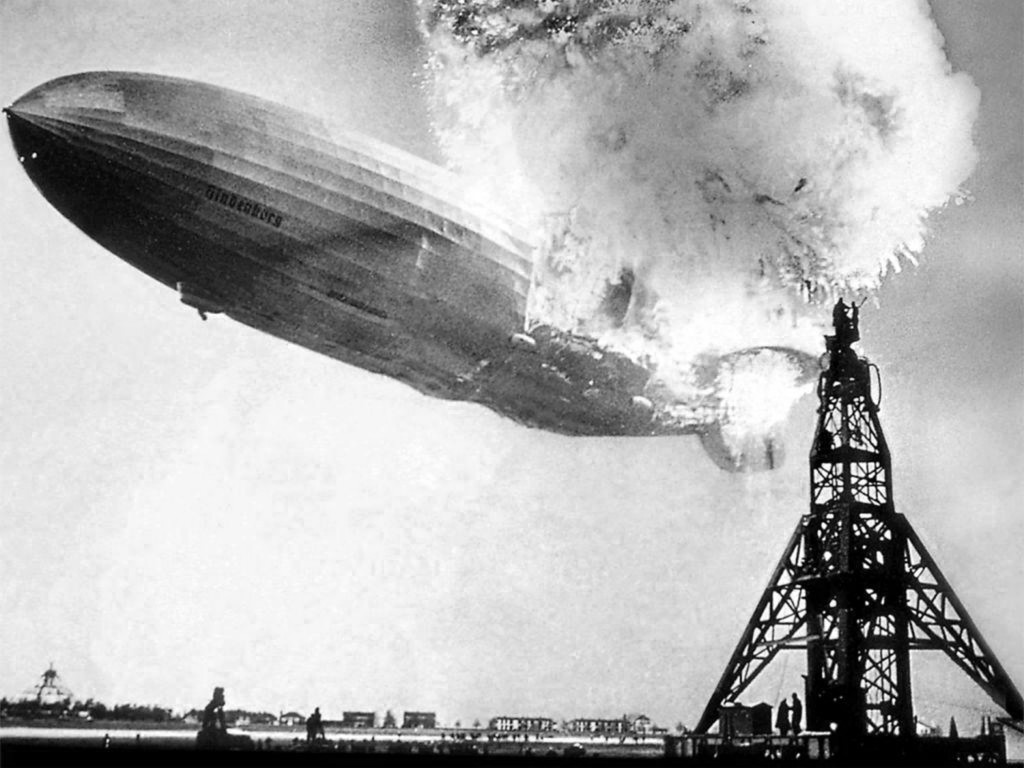}
\caption{Hydrogen fire destroys Hindenburg airship on May\,6,\,1937. Such disasters could be preventable if the airships were equipped with sensors capable of detecting low concentrations of leaking hydrogen gas. From \url{https://www.airships.net/hindenburg/disaster/}.\label{Fig1}}
\end{figure}   

However, as with other clean energy technologies, numerous technical challenges have thus far prevented {\Ivan a large-scale use of hydrogen in the mainstream technological processes and practices}. These include the expense of its mass production and long-term storage, because hydrogen rarely occurs naturally as a gas on Earth, and safety concerns due to the high reactivity of hydrogen fuel with environmental oxygen in the air.

The problem of safety of hydrogen-based technologies is especially pressing because hydrogen is flammable over a very wide range of concentrations in air (4\%–75\%) and it is explosive over a wide range of concentrations (15\%–59\%) at a standard atmospheric temperature \cite{Hub14}. In other words, when ignited in an enclosed space a hydrogen leak will most likely lead to an explosion, not a mere flame. This makes the use of hydrogen particularly dangerous in enclosed areas such as tunnels and underground parking. For such safety reasons the development of hydrogen airship technologies was completely abandoned in the 1930s, after several deadly crashes culminated by the infamous Hindenburg Disaster that was caused by an electrostatic discharge that ignited leaking hydrogen (Fig.~\ref{Fig1}). Obviously, such a disaster could be prevented if appropriate technologies, including sensors sensitive to low concentrations of leaking hydrogen, were available.

Thus, to benefit from the hydrogen’s ability to act as a clean-energy carrier and to transition to a hydrogen economy, there is an urgent need in efficient and low-cost hydrogen gas sensors. Such sensors are needed anywhere where hydrogen gas is produced or consumed, in particular (i)~to ensure safety of cars and hydrogen gas fuelling stations and (ii)~to measure concentration of hydrogen gas inside fuel cells. {\Ivan (It is noteworthy that hydrogen also acts as an indirect greenhouse gas and, therefore, preventing its leakage is important not exclusively only due to the safety reasons.)} The safety sensors must be capable of sensing even weak leakages of hydrogen gas and therefore have to be susceptible to tiny gas concentrations (in the ppm range). Significantly, the response time of hydrogen sensors must be less than one second \cite{Hub14} to ensure that the concentration of leaking hydrogen does not exceed a dangerous level. In turn, the hydrogen concentration meters have to demonstrate high concentration resolution in the concentration ranges above 20\% not losing their measuring ability near the 100\% concentration mark \cite{Hub14}. 

These stringent requirements challenge the existing hydrogen sensor technologies \cite{Hub14}. In fact, no one sensor has thus far demonstrated optimal performance under conditions expected to be encountered in a practical hydrogen technology situation \cite{Kor09, Hub11, Hub14, Cha19}. Furthermore, only a small number of existing solid-state device-based hydrogen concentration meter concepts have demonstrate a non-vanishing concentration resolution above 20\% of hydrogen in the environment \cite{Hub14}.
\begin{figure}[t]
\includegraphics[width=7.5 cm]{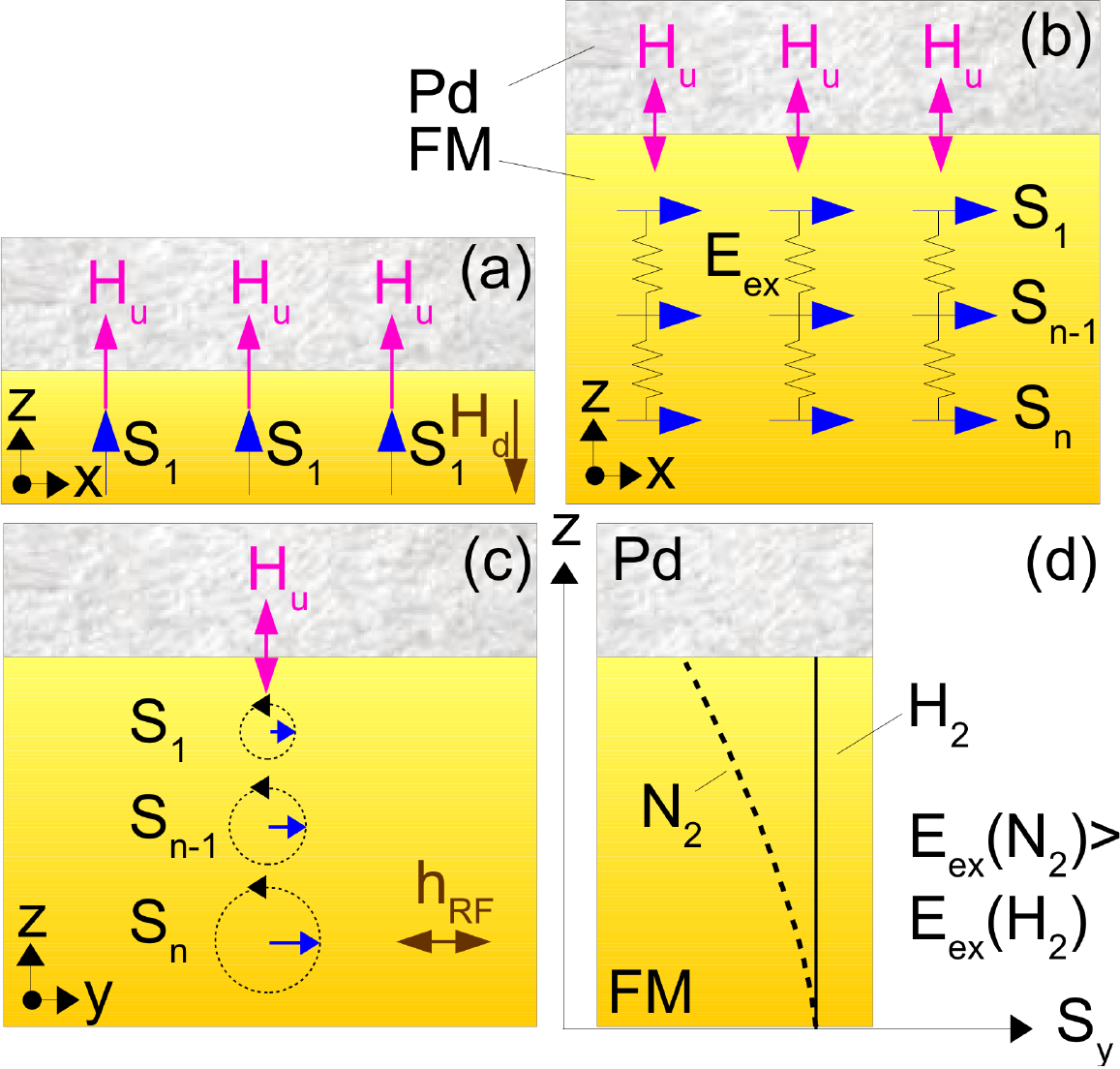}
\caption{Simplified pictures of the magnetisation ground state of a continuous film in the presence of PMA. (\textbf{a})~Pd/FM-layer film with a {\OKB single atomic} FM layer. All electron spins ${\bf S}_i$ point perpendicular to plane because $|{\bf H}_u|>|{\bf H}_d|$, an arrangement called {\OKB the} perpendicular magnetic medium {\OKB (see the main text for the meaning of physical parameters used in this figure).} (\textbf{b}){\OKB~An $n$-atomic-layers} thick film is an effectively in-plane magnetised medium because $|{\bf H}_u|/n < |{\bf H}_d|$. (\textbf{c})~Low-energy excitation on top of the ground state in Panel~(b) in the form of a {\OKB resonance precession of the spins known as FMR} ($h_{RF}$ is the applied microwave magnetic field). (\textbf{d})~Amplitude of magnetisation precession as a function of $z$ for the H$_2$ and N$_2$ gas atmospheres.\label{Fig2}}
\end{figure}   

It is noteworthy that, in general, hydrogen sensors technologies are relatively mature since such sensors have been used for decades in various industrial environments such as the petroleum, food, chemical and aerospace industries (see \cite{Hub14} for a review). However, those sensors have been designed to operate in specific environments that, strictly speaking, are very different from those of fuel cells and relevant technologies. This urges the development of new hydrogen sensors designed to meet the demands of the emergent hydrogen economy. The literature summarising and discussing the recent progress in this rapidly growing research field is abundant (for a review see, e.g., \cite{Gup08, Kor09, Hub11, Hub14, Cha19, Shi21}), and thus the reader is referred to those articles for further details. However, despite significant advances made in this area, most of the hydrogen gas sensor concepts proposed thus far have several major drawbacks, including poor sensitivity, complex detection systems, slow response time, high power consumption and potential flammability issues \cite{Hub11, Gup12}. Consequently, the search remains open for novel technologies that would resolve the aforementioned technological challenges. 

This review article is devoted to a new kind of hydrogen sensors that hold great promise to fill the niche of hydrogen sensors capable of supporting the transition to a hydrogen economy. Recently, a novel, ultra-low fire-hazard, contactless and simple method for hydrogen gas sensing based on a magnetic (magneto-electronic) approach was proposed \cite{Cha13}. This kind of sensors exploits changes in magnetic properties of thin magnetic films made of alternating layers of cobalt (Co) and palladium (Pd) or a single layer of an alloy of Co or iron (Fe) with Pd \cite{Cha13, Lue16, Lue19}. This approach has been tested in a range of hydrogen concentrations from 0.06\% to 100\% at atmospheric pressure and showed highly promising results: the prototypes of magneto-electronic hydrogen sensors have been able to detect the presence of hydrogen gas all across this range of concentrations. This success warranted the writing of this review article with the main objective to inform the academic physics, chemistry and engineering communities as well as industry researchers about the recent developments in the field magneto-electronic hydrogen sensors, thereby facilitating the translation of research results into policy and practice. In particular, we present the physical foundations of magneto-electronic hydrogen sensors and we critically overview their advantages and disadvantages for applications in the vital areas of the safety of fuel cell cars and hydrogen gas fuelling, including hydrogen concentration measurement inside fuel cells.
 
\section{Physical foundations of magneto-electronic hydrogen gas sensors}
We start with a brief overview of the general physical phenomena underlying the operation of magneto-electronic hydrogen sensors. The reader interested in a comprehensive discussion of these effects can refer to the following original, review and tutorial articles: \cite{Cha88, Tse02, Cha07, And11, Bra12, Mak15, Zha15, Pri98, Zvezdin, Arm13, Mak15_review, Mak16_1}. Further discussions of particular physical phenomena in the context of magneto-electronic hydrogen sensors will be provided on an {\it ad hoc} basis in the text below, {\Ivan where we will often refer to the effect of perpendicular magnetic anisotropy (PMA) and the concept of perpendicular magnetic media illustrated in Fig.~\ref{Fig2} and where we will also use the following notation} {\OKB: ${\bf S}_i$ denotes the electron spins, ${\bf H}_u$ is the effective field of PMA, $|{\bf H}_d|$ is the perpendicular-to-plane demagnetising field that scales as {\OKB the volume density of spins} and does not depend on the number of {\OKB atomic} layers $n$. Parameter $E_{ex}$ denotes the {\OKB inhomogeneous} exchange energy that tends to align the spins parallel to each other ($E_{ex} = 0$ for co-aligned spins as shown schematically by springs connecting the spins). In this simplified physical picture, different {\Ivan  strengths} of ${\bf H}_u$ for the H$_2$ and N$_2$ gas atmospheres translate into a difference in the {\OKB inhomogeneous} exchange energy contribution $E_{ex} \sim (d{\bf S}_i/dz)^2$ to the energy of {\OKB excitation above this ground state} and thus to a difference in the respective ferromagnetic resonance (FMR) frequencies}.

Magneto-electronics \cite{Pri98, Joh04, TechGuide, Mak16} is a rapidly growing sub-field of the broader areas of nanoscience and nanotechnology. Significant progress has been achieved in this area during the last decades, resulting in new technologies for information storage and signal processing \cite{Pri98, Mak16, Cai18}. One of the most important landmarks in this field was the discovery of perpendicular magnetic anisotropy (PMA) \cite{Cha88, Wel94}. This type of uniaxial anisotropy exists at the interface of a ferromagnetic metallic (FM) layer with several non-magnetic (NM) metals, in the first instance with platinum (Pt) and palladium (Pd) \cite{Wel94}. In such structures, anisotropy is very strong and has its axis directed perpendicular to the film plane \cite{Tsy12}. Therefore, to take advantage of this interface effect, materials should be from several angstroms to several nanometres thick. Consequently, approaches from cutting-edge nanotechnology techniques are employed to fabricate and study these materials.

{\Ivan For instance, the use of magnetic films with PMA has revolutionised the technology for magnetic hard drives resulting in the large capacity of hard drives for desktop and laptop computers that is so important today. These advances have also paved the way for new applications of magnetic multilayers and nanostructures in adjacent fields of research and engineering \cite{Cha07}. However, the key materials used in these technologies and, therefore, the key fundamental physical effects underpinning their functionality, have also played an important role in other technologies that have been known for more than 150\,years. Indeed, Pd--the key material of the hydrogen sensors discussed in this review article--has also been the key model material in the fundamental studies of the interaction of hydrogen with metals \cite{Ale78, Fuk05, Gol17}. Despite a long history of research on this topic, only in the 1970s it was shown that a controllable and reversible change in the action of hydrogen on Pd results in significant changes in the structure and properties of the latter \cite{Gol73, Gol81, Fuk05, Gol17}. Some of these important findings are also relevant to the forthcoming discussion below.}  

\subsection{Perpendicular magnetic anisotropy\label{sec:2.1}}
Spins of localised $d$-electrons are carriers of magnetism in ferromagnetic metals. The density of these electrons is large and this allows one to consider a macroscopic magnetisation vector that represents the vector sum of the spins per unit volume{\OKB ($\mu_B \sum_i {\bf S}_{i}/V_i$, where $\mu_B$ is the Bohr magneton, see Fig.~\ref{Fig2})}. In thin magnetic films, the magnetisation vector lies naturally in the film plane because of a very large out-of-plane demagnetising field ${\bf H}_d$ (more than 1.8\,T for Co films). In a typical Co/NM bilayer or multilayer film, an effective magnetic field of {\OKB Perpendicular Magnetic Anisotropy (PMA) }, ${\bf H}_u$, is induced at the interface of a Co nanolayer with a Pt (“heavy metal”) group NM nanolayer \cite{Gra86, Dra87}. For example, PMA {\OKB has} been observed at an interface of a Pd layer with an FM metal such as Co, Fe, Ni-Fe and their alloys \cite{Eng91}. While interfacial stress and symmetry breaking are among the key factors leading to PMA \cite{Cha88}, the presence of ${\bf H}_u$ makes the perpendicular-to-plane orientation of the spins of the upper-most atomic layer [so-called interface spins, see Fig.~\ref{Fig2}(b)] energetically preferable. However, in FM materials the interface spins are coupled to bulk spins by {\OKB a} strong exchange interaction that tends to co-align them parallel to bulk spins. As a result, the overall impact of ${\bf H}_u$ on the magnetic properties of the material scales is $1/t$, where $t$ is the thickness of the FM layer \cite{Cha88}. For thinner FM layers ($t$ = 0.3--1\,nm), ${\bf H}_{u}/t > {\bf H}_d$ and the magnetisation vector is oriented out-of-plane [Fig.~\ref{Fig2}(a)]. For thicker films, ${\bf H}_u/t < {\bf H}_d$ and hence the magnetisation vector flips into the film plane [Fig.\,\ref{Fig2}(b)].

However, using ultra-thin films is not always convenient from the technology standpoint. Hence, superlattices that represent many repeats of NM/FM bilayer blocks have been suggested as a valid approach to keep the magnetisation vector out-of-plane for samples with a total thickness of the order of tens of nanometres \cite{Eng91}. Several physical responses of magnetic materials scale with the volume $V$ (and thus with the film thickness $t$), including the total magnetic moment {\OKB ($\mu_B \sum_i {\bf S}_i$)} and the microwave power absorbed in {\OKB Ferromagnetic Resonance (FMR)} measurements (Sec.~\ref{sec:2.2}). For example, the effect of interface PMA on a bulk property such as the FMR frequency scales as $1/t$. Hence, for an about 40-nm-thick single-layer Co film the effect of PMA on the FMR frequency becomes negligibly small and therefore no hydrogen-induced frequency shift is observed (i.e. there is no sensitivity to hydrogen). However, using superlattices exhibiting the PMA effect allows observing easily measurable physical responses to which the interface effect of PMA makes a dominant contribution.
\begin{figure}[t]
\includegraphics[width=6.0 cm]{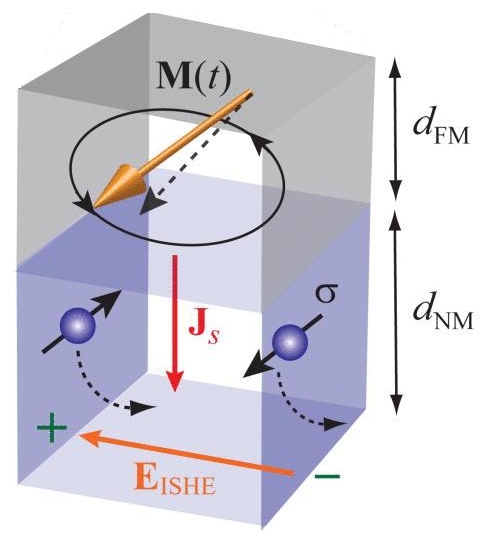}
\caption{Illustration of the spin pumping (SP) and inverse spin Hall effect (iSHE) effects. {\OKB Parameters ${\bf M}(t)$, ${\bf E}_{iSHE}$, ${\bf J}_s$, $\sigma$, $d_{FM}$ and $d_{NM}$ represent {\OKB the} dynamic magnetisation {\OKB vector of the ferromagnetic metallic (FM) layer}, an electric field induced due to the iSHE {\OKB in the non-magnetic metallic (NM) layer}, the spatial direction of the spin current, spin polarisation of the spin current and thickness of the FM and NM layers, respectively. The black dotted arrow in the NM layer {\Ivan depicts} {\OKB the conduction-electron} motion induced by spin-orbit interaction and responsible for the iSHE.} {\Ivan Reprinted from \cite{Yos12}, with the permission of AIP Publishing.}\label{Fig3}}
\end{figure}   

\subsection{Ferromagnetic resonance \label{sec:2.2}}
{\OKB The} ferromagnetic resonance (FMR) is the effect of spin precession of the macroscopic vector of magnetisation in an external magnetic field (for a review see, e.g., \cite{Mak15}). In FMR, the spin precession is driven by an external source of microwave power [$h_{RF}$ in Fig.~\ref{Fig2}(c)]. The unique characteristics of FMR has been used in microwave technology for many decades \cite{Mak15}. For example, the frequency of spin precession is determined by the material parameters, the applied magnetic field and geometry of the sample. The onset of FMR is registered as an increase in absorption of microwave power by the sample when the microwave frequency of the external source matches the FMR frequency. 

Differential FMR traces are observed when an additional small ac magnetic field is applied to the sample parallel to the static magnetic field, the output signal is rectified using a microwave diode and fed into a lock-in amplifier referenced by the same ac signal \cite{Mak15}. As a result, the recorded resonance line has the shape of the first derivative of a Lorentzian shape. The resonance absorption can be observed by gradually changing the frequency $f$ of the microwave signal and fixing the external magnetic dc field $H$ applied to a sample, or vice versa by fixing $f$ and varying $H$ \cite{Mak15}. {\Ivan The latter (i.e.~the applied-field-resolved) method is used more widely because of its convenience and the absence of artefacts in the collected data that can occur due to microwave-frequency dependence of parameters of employed microwave gear.} 

In \cite{Cha13} it was found that absorption of H$_2$ shifts the FMR frequency of a Pd[10\,nm]/Co[5\,nm] film by 300\,MHz at 18\,GHz. The results of that study have enabled using the FMR response as a physical effect underpinning the development of novel Pd/Co-based hydrogen gas sensors.

The FMR frequency is shifted because the spins at the interface cannot precess as freely as in the bulk of the layer due to ${\bf H}_u$ (so-called interface spin pinning, see Fig.~\ref{Fig2}(c), where, for clarity, we neglected the ellipticity of precession). Through exchange coupling, the effect of interface pinning is propagated across the whole thickness of the FM layer, resulting in a spatially non-uniform precession amplitude profile {\OKB $S_{y}(z)$ [Fig.~\ref{Fig2}(d)]. The non-uniformity of $S_{y}(z)$ leads to a non-vanishing contribution of $E_{ex}$ to the FMR frequency. The larger ${\bf H}_u$, the larger the uniformity of $S_{y}(z)$} and the larger $E_{ex}$ [Fig.~\ref{Fig2}(d)]. As a result, the FMR frequency becomes a function of ${\bf H}_u$. This was illustrated by the pioneering experiment in \cite{Cha13} involving swapping between N$_2$ and H$_2$ gas atmospheres in the chamber (see Sec.~\ref{sec:3.1} for details) while keeping the frequency of the microwave field constant.

{\Ivan Note that the non-uniformity of the profile of the precession amplitude and changes in the FMR frequency due to the respective increase in the inhomogeneous-exchange contribution to magnetic energy are indistinguishable from an effect induced by a thickness-uniform effective field of the bulk PMA. Subsequently, experimentalists may use the strength of the effective bulk PMA field to quantify this intrinsically interfacial effect. Importantly, the resulting effective bulk PMA field scales as the inverse thickness of the FM layer that forms an interface with NM layers where the interface PMA exists, and its hyperbolic dependence on the magnetic layer thickness is often used to distinguish an interface PMA from a bulk PMA.}

\subsection{Spin pumping and interface clearing\label{sec:2.3}}
Spin-pumping occurs in a bilayer film comprising an FM layer and an NM layer in electrical and/or exchange contact with the FM layer \cite{Tse02, Tak16}. If magnetisation precession is excited in the FM layer by a microwave source, an angular momentum is transferred from microwave photons to magnons--the quanta of magnetisation precession--in the ferromagnetic layer. This angular momentum leaks into the NM layer as a spin flow ${\bf J}_s$ from the FM to the NM layer (Fig.~\ref{Fig3}, see \cite{Yos12}). That is, magnetisation precession acts as a spin pump that transfers angular momentum from the FM layer into the NM layer. From the macroscopic standpoint, this loss of angular momentum by the FM layer acts as additional damping of magnetisation precession and is often registered as a resonance linewidth broadening in FMR experiments.
 
In \cite{Cha13}, it was found that absorption of H$_2$ by Pd decreases the FMR linewidth. This decrease is readily detectable {\OKB--}15\% at 18\,GHz. One potential contribution to this effect is reduction in spin pumping. The fact {\OKB that} this contribution exists was recently evidenced in \cite{Wat20} by measuring {\OKB an} Inverse Spin Hall voltage for a Pd/Co bilayer film (see below). Significantly, the FMR response scales as the inverse damping and therefore the decrease in magnetic damping also results in a 15\% increase in the microwave absorption amplitude for the maximum of the resonance line. As a result, a decrease in FMR linewidth can also be used as a mechanism for reading the state of magneto-electronic hydrogen gas sensors.

{\OKB The interface clearing effect is another physical mechanism that can potentially result in a decrease in the FMR linewidth in the presence of hydrogen gas \cite{End05}. Evidence supporting this assumption has been produced in a recent experiment involving Pd/Y/NiFe trilayer films \cite{Wei21}. The interface clearing effect results in a reversible improvement of structural quality of an interface formed between an NM metal capable of absorbing hydrogen (Pd or Y) and an FM metal. It has been speculated in \cite{Wei21} that in the presence of hydrogen the atoms of the interface become more mobile and therefore can rearrange themselves such that the interface becomes more regular. {\OKB This} reduces scattering of the FMR precession mode from the interface inhomogeneities {\OKB, which yields a narrower FMR line and a more pronounced FMR peak}. Significantly, this effect is fully reversible and, when hydrogen is removed from the material, the interface regains its original behaviour.} 

\subsection{Inverse Spin Hall Effect\label{sec:2.4}}
Exploiting the Inverse Spin Hall effect (iSHE) opens up opportunities for cheap and fire-safe reading of the state of a magneto-electronic hydrogen sensor. In an FMR experiment, iSHE manifests itself as transformation of a signal of microwave magnetisation precession into a dc voltage \cite{Val06, Sai06, Kim07, San11}) and thus represents an excellent tool for electrical detection of an FMR signal. In an NM layer, ${\bf J}_s$ produced by SP is carried by spin-polarised conduction electrons (Fig.~\ref{Fig3}) that diffuse {\OKB from the FM layer through} the interface and are scattered by ions of the crystal lattice of the NM material (dashed lines) {\OKB in the plane} of the NM layer. (There is also a back flow of electrons towards the interface to ensure electro-neutrality of the interface.) Due to the spin-orbit interaction that is particularly strong in heavy metals, {\OKB in the presence of an external constant magnetic field,} the scattering results in an effect analogous to the Lorenz force exerted by a magnetic field on conduction electrons: a {\OKB spin-polarised conduction} electron is deflected in the layer plane in a direction that is dependent upon the directions of its spin and its instantaneous velocity. This creates a dc voltage $E_{iSHE}$ in the plane of the NM layer. At room temperature, the value of $E_{iSHE}$ attains tens of mV across a length of approximately 1\,cm. This value is large enough to ensure reliable detection of this voltage, but at the same time is small enough not to represent a fire hazard in a hydrogen-containing atmosphere. Therefore, iSHE can be used for hazard-free reading of the state of a Pd/FM-based hydrogen gas sensors, which has recently been confirmed experimentally \cite{Wat20}.

It is noteworthy that there exists a fundamental difference between the iSHE-based sensing scheme and a popular concept of a gas sensor based on measurement of the electrical resistivity of a single Pd layer. To measure the resistivity, one must apply a dc voltage in an H$_2$ atmosphere. A fault in a voltage-supplying circuit may result in a voltage level that is much higher than the nominal one and hence application of a dc voltage to a sensor poses a potential fire hazard. However, no dc voltage needs to be applied from outside to a sensor based on iSHE since dc voltage is induced naturally in a heavy metal by virtue of the iSHE. This voltage is so small that a fire hazard is highly unlikely.

{\OKB \subsection{Magneto-optical Kerr effect and anomalous Hall effect}\label{sec:2.5_MOKE_Hall}
The burgeoning interests in thin films and nanotechnologies have motivated the development of a number of innovative spectroscopy and microscopy tools, some of which exploit the magneto-optical effects \cite{Zvezdin, Arm13, Mak15_review, Mak16_1} to investigate physical properties of magnetic materials \cite{All03, Bar08, Sol17, Cha19_Taiwan}. For example, the magneto-optical Kerr effect (MOKE) represents a change in the polarisation and intensity of light that is reflected from the surface of a magnetic material. Similarly to the Faraday effect \cite{Zvezdin, Arm13}, the MOKE originates from the off-diagonal components of the dielectric permittivity tensor of the investigated magnetised material \cite{Zvezdin}. However, while the Faraday effect works in transmission and thus occurs only in optically transparent materials, the observation of the MOKE is possible mostly in highly optically-reflecting samples. Due of this property, the MOKE has been found to be especially suitable for studying magnetism of metals because metallic surfaces are usually characterised by strong specular reflection of light.

Depending on the direction of sample magnetisation with respect to its surface and the plane of incidence of light, the MOKE can be observed in the polar, longitudinal and transverse configurations \cite{Zvezdin, Arm13}. All three MOKE configurations have been used to characterise magnetic materials. Significantly, the magneto-optical constants describing the contribution of MOKE to the off-diagonal components of the tensor of dielectric permittivity of the material depend on the value of sample magnetisation and, as a result, the angle of polarisation rotation for the light reflected from the surface of the sample is also a function of magnetisation. This dependence enables one to measure hysteresis loops for highly reflecting magnetic samples using the MOKE. Note that the scale of the vertical axis of hysteresis curves obtained using this method is arbitrary. {\OKB Hence,} to convert a MOKE signal into the sample magnetisation one needs to know the values of magneto-optical constants for the material under study. However, such data are often unavailable in the literature, which is especially the case of metal alloys with arbitrary compositions. {\OKB In addition,} the magneto-optical material parameters may change during the experiment, for instance, as a result of absorption of H$_2$ gas by the metallic sample \cite{Liu11, Tit12, Str13, Tit14, Nug19, Hek20}.  

The anomalous Hall effect \cite{Ger02, Ger08, Nag10, Hir20} belongs to the family of Hall effects in magnetic substances. It is well-known that the ordinary Hall effect occurs in a conductor in the presence of a magnetic field and an electric current flowing perpendicular to it, and it is observed as an electric voltage induced in the direction that is perpendicular to both the electric current and magnetic field \cite{Hur72}. {\OKB I}n FM materials, there is an additional contribution to the Hall voltage known as the anomalous Hall effect or the extraordinary Hall effect \cite{Kar54, Ger02, Ger08, Nag10, ZheESE_19, Hir20}. This contribution to the Hall voltage depends on the material magnetisation value and in some materials it can be much larger than the contribution of the ordinary Hall effect \cite{Ger02, Nag10}. Since the anomalous Hall effect {\OKB may} originate from spin-dependent scattering of the charge carriers, one {\OKB may} expect that a change in the magnetic state and/or electrical conductivity of the sample under study would result in a change in the Hall voltage \cite{Ger02, Ger08, Nag10, ZheESE_19, Hir20}.}

\subsection{Mechanism of hydrogen gas absorption by palladium\label{sec:2.5}}
Pd is one of the most used materials in hydrogen gas sensing because it features reversible absorption of H$_2$, leading to the formation of palladium hydride \cite{But11}. During absorption, certain physical properties of the material change and these changes are used to detect the presence of H$_2$ and its concentration. In particular, the atomic lattice parameter of Pd can increase by up to 3\% and its resistivity by up to 80\%. {\Ivan Pd is highly selective to H$_2$ absorption and it exhibits a much lower sensitivity to other gases such as CO, Cl$_2$, SO$_2$, H$_2$S, NO$_x$ and hydrocarbons \cite{Gup12} (but these gases can still poison the sensor and therefore special protection measures may need to be taken \cite{Hub11})}. These properties make this material unique for applications in hydrogen gas sensing. 

When a hydrogen molecule approaches the surface of palladium metal, the distance between the two hydrogen atoms increases due to a strong interaction between the atoms of Pd and H \cite{Bar04}. Through this process the strength of the H-H bond is weakened \cite{Bar04} and, as a result, the hydrogen molecule is dissociated into two unbonded hydrogen atoms. The atomic radius of a hydrogen atom is small (around 0.53\,{\AA} in the ground state) \cite{Lin16} and therefore hydrogen atoms can easily diffuse into materials.

\begin{figure}[t]
\includegraphics[width=7.0 cm]{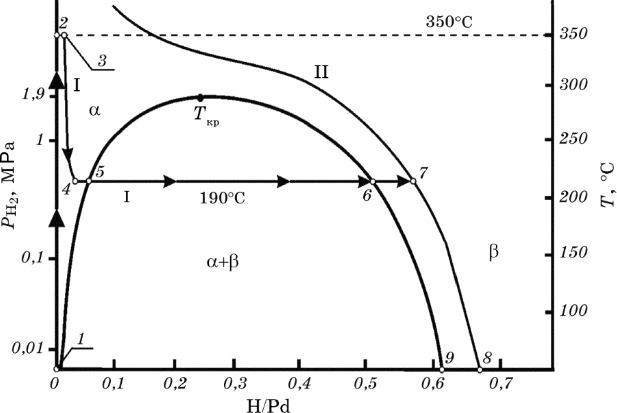}
\caption{{\Ivan Equilibrium phase diagram of a Pd--H system and the pathway of palladium hydrogenation to the hydride state (points from 1 to 8) showing the $\alpha \to \beta$ hydride phase transformation. I and II denote the isobars for 0.29\,MPa and 2.3\,MPa. Reprinted from \cite{Gol17}. The interested reader is also refered to \cite{Fuk05}.}\label{Fig4}}
\end{figure}

Figure~\ref{Fig4} shows a phase diagram of palladium hydride. Hydrogen atoms absorbed in Pd may form $\alpha$- and $\beta$-palladium hydride phases depending on the hydrogen concentration inside the metal. Hydrogen atoms mainly occupy the interstitial sites of the crystalline lattice for the both phases \cite{Ohn14}. At low hydrogen concentrations, the solid solution, $\alpha$-phase, is formed. While the hydrogen concentration increases, the metal hydride, $\beta$-palladium hydride, is formed. The hydrogen molecules are dissociated into atoms and firstly adsorbed at the surface of palladium by the high-symmetry hollow chemisorption sites, and then filling the octahedral interstitial sites in the first subsurface layer and finally diffuse and penetrate the interstitial sites inside the bulk \cite{Ohn14}. At room temperature, the pure $\alpha$-phase palladium hydride corresponds to a stoichiometry $x < 0.017$ while pure $\beta$-phase is realised for $x > 0.58$ (where $x$ is the stoichiometry of PdH$_x$). The intermediate values of $x$ correspond to mixtures of both phases. At lower hydrogen concentration (less than PdH$_{0.017}$, i.e.~$\alpha$-phase), the lattice of palladium slightly expands from 3.889\,{\AA} to 3.895\,{\AA}. For higher concentrations, the $\beta$-phase Pd-H will form and the palladium lattice size will increase to 4.025\,{\AA} \cite{Man94, Oll13}. When Pd represents a thin film clamped to a substrate, it cannot expand horizontally, but only vertically. This induces significant internal stresses in Pd films. Pd layers with thicknesses on the order of 10\,nm can easily withstand these stresses, but ones with thicknesses in the range above 30\,nm develop blisters, wrinkles and peel off the substrate \cite{Lee10}.   

{\OKB Furthermore, the results from \cite{Har17} demonstrate that hydrogenation of Pd thin films results in two stages of lattice expansion depending on the concentration of hydrogen gas. At low, up to 2\% of hydrogen gas concentrations, the lattice constant increases only in the out-of-plane direction by approximately 1\%. This expansion is completely reversible, i.e.~the thickness of the Pd layer returns to its original value after desorption. However, in the second stage the lattice constant grows by up to 4\% in both out-of-plane and in-plane directions despite substrate clamping. Therefore, because of the lateral expansion, these changes are irreversible and they cause structural changes to the Pd lattice. We note that a reversible vertical expansion was also observed in thin Pd layers grown on top of Co layers \cite{Cal16}.} 

\subsection{Pd in electronics, magneto-optics and plasmonics \label{sec:2.6}}
One of the important technological applications of Pd is electronics, where it is used in multilayer ceramic capacitors as well as for component and connector plating in consumer electronics \cite{Mro98}. Pd also has excellent optical properties originating from its ability to support localised and surface plasmonic waves \cite{Lan06}. While surface plasmons are waves that propagate along a metal-dielectric interface, a localised surface plasmon is the result of a tight confinement of a surface plasmon by a metallic nanoparticle that is smaller than the wavelength of the incident light. Metallic nanoparticles at or near their plasmonic resonance can generate highly localised electric field intensities. Varieties of nanoparticles and their constellations (e.g.,~nanoantennas \cite{Mak16_1}) have been used to enhance the local field and thus improve light-matter interaction that is so important for a number of sensing applications.

Subsequently, nanostructures made of Pd or combining Pd with other plasmonic metals such as gold and silver have been used in hydrogen gas sensors \cite{Liu11, Wad14, Tit14, Nug19, Che_arxiv}. Similarly to other types of gas sensors, gas detection by plasmonic sensors is achieved by measuring a shift in the resonance peak in the spectrum of light scattered by a nanostructure exposed to gas. Rigorous numerical simulations helped to establish that a hydrogen-induced spectral shift originates from two competing physical effects: a small blue-shift caused by changes in the dielectric function of Pd and a much stronger red-shift due to an expansion of the lattice of Pd \cite{Tit12}.

It is also important to mention application of Pd in magneto-optics, which gives rise to an important class of magneto-optical hydrogen gas sensors, where, for example, the magneto-optical Kerr effect (MOKE) is employed \cite{Lin12, Lin13_1, Lia17_Taiwan, Cha18_Taiwan, Cha19_Taiwan} (see Sec.~\ref{sec:MOKE} for more details). It is noteworthy that the strength of the MOKE response can be enhanced using plasmonics properties of Pd and FM metals used in hydrogen sensitive multilayers \cite{Mak15_review}.
\begin{figure}[t]
\includegraphics[width=10.5 cm]{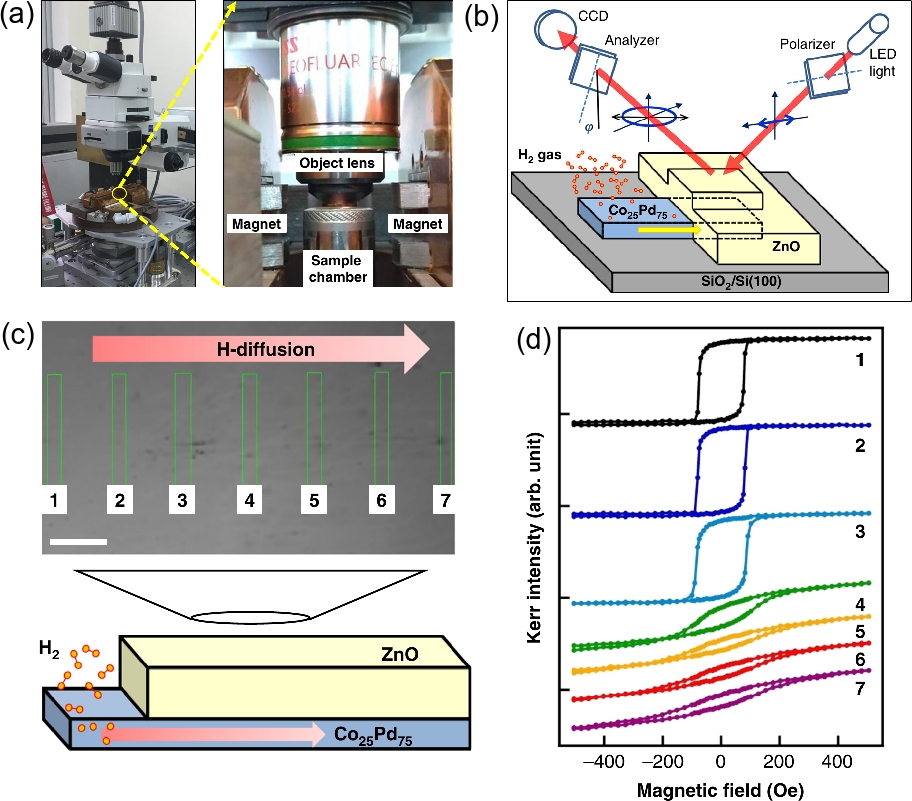}
\caption{\textbf{Top row}:~Magneto-optical Kerr effect (MOKE) setup. \textbf{(a)}~General view and a close-up showing the objective lens, poles of the magnet and test chamber, where the sample is located. \textbf{(b)}~Schematic of the sample and the MOKE measurement principle. \textbf{Bottom row}:~The experiment, where the magnetic properties of the sample due to the absorption of hydrogen were investigated. \textbf{(c)}~Optical microscope image of the sample with the arrow indicating the direction of hydrogen diffusion (the scale bar is 20\,$\mu$m). \textbf{(d)}~MOKE hysteresis loops measured at the areas of the sample indicated by the rectangles in Panel~(c). Reprinted with permission from \cite{Cha19_Taiwan}.\label{MOKE_setup}}
\end{figure}   

{\OKB Not only multilayers but also alloys of Pd with Co, Fe and Ni have been intensively investigated for applications in hydrogen sensing \cite{Lin12, Lee12, Lin13_1, Lin16, Win15, Ger17, Hsu17, Lia17_Taiwan, Aka19, Wan18_Taiwan, Cha18_Taiwan, Cha20_Taiwan_alloy} using a combination of different methods that included: MOKE spectroscopy, high-resolution transmission electron microscopy, enhanced x-ray magnetic circular dichroism technique, atomic force microscopy, in-operando x-ray and neutron spectroscopies as well FMR spectroscopy. It is worth noting here that using Pd alloys in FMR-based hydrogen detection one faces technological challenges such as an increase in the linewidth of the FMR peak and a concomitant decrease in the peak amplitude, which is well-known fact established in FMR measurements of samples made of impure FM metals \cite{Mak15}.} 

\subsection{\OKB Early studies of the effect of hydrogen on Pd-based magnetic materials and theoretical explanation of the effect}\label{sec:2.7}
{\OKB Most of investigations of the interface PMA from the literature have been conducted using Pt as the heavy metal layer. This is because Pt-based multilayers demonstrate a stronger interface PMA than any other metal of the Pt group.} Another important element from the same group--Pd--has attracted much less attention. {\OKB However}, some of the pioneering experiments on PMA in multilayers used Pd layers alongside Pt ones \cite{Wel94} demonstrating an efficiency comparable with that of Pt/FM systems \cite{And10}. {\OKB Furthermore, the inverse spin Hall effect also exists for Pd/FM bilayers \cite{Har07}.}

Pd/FM bilayers develop PMA \cite{Eng91_1, Law07} and are characterised by a strong SP effect \cite{For05, Sha12}. Some studies have demonstrated changes in the strength of PMA caused by the loading of Pd/Co superlattices with H$_2$. For example, the pioneering work \cite{Oka02} showed an initial increase in the strength of PMA due to absorption of H$_2$ and then a decrease for larger partial pressures of H$_2$. However, the experiments in \cite{Oka02} were carried out for partial pressures 1.3--5.2\,atm. The works \cite{Mun11, Mun12} seem to be the first detailed investigations of changes in the strength of PMA induced by absorption of H$_2$. In particular, these papers report combined structural studies of the interface between Pd and Co layers with magnetic studies of PMA and magnetisation, and they demonstrate a decrease in both parameters as a function of H$_2$ absorption. The experiments in \cite{Mun11, Mun12} were conducted at 1\,atm and the atmosphere was pure H$_2$.

{\OKB The mechanism of the effect of hydrogen on the Pd-Co chemical bond and the resulting change in magnetism was explained in \cite{Kly20} using Density Functional Theory (DFT) calculations, where a Pd/Co interface was studied as a model system. The DFT calculations demonstrated that hydrogen insertion at the Pd/Co interface changes the interface PMA by modifying the electronic structure of the interface. It was found that the accumulation of hydrogen at the Pd/Co interface affects the hybridisation between neighbouring Co and Pd layers, leading to a decrease of the perpendicular anisotropy component, and that eventually it may change the net magnetic anisotropy to the in-plane state. Interestingly, it was also found that when hydrogen penetrates into the interior of Co, it has the opposite effect of promoting the PMA state. These changes are governed by competing contributions of the $d_{xy}$; $d_{x^2+y^2}$ and $3d{z^2}$; and $3d_{zy}$ states that are mainly responsible for the perpendicular and the in-plane magnetocrystalline anisotropy, respectively.}

This theoretical finding is in agreement with an earlier experiment, where the electronic origin of the change to the interface PMA induced by the presence of hydrogen was demonstrated by excluding other possible options \cite{Lue17}. We will return to this experimental result below.
\begin{figure}[t]
\includegraphics[width=13.5 cm]{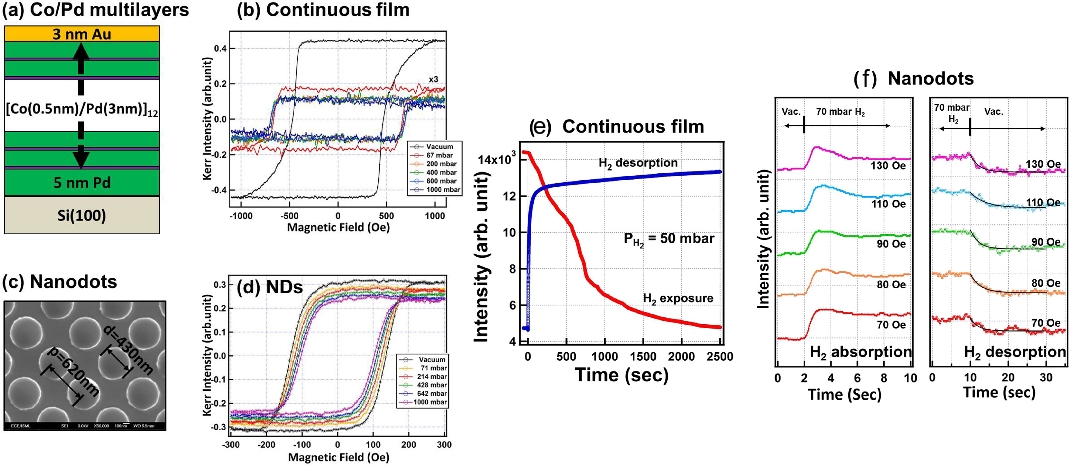}
\caption{ \textbf{(a)}~Schematic of the investigated Pd/Co multilayer continuous film and \textbf{(b)}~its MOKE hysteresis loops measured in vacuum and various at H$_2$ gas pressure levels. \textbf{(c)}~Representative scanning electron microscopy photograph of the nanodots sample with the typical rod diameter and inter-dot spacing. \textbf{(c)}~MOKE hysteresis loops of the nanodots sample. \textbf{(e, f)}~Real-time monitored optical response of the continuous film sample and nanodots. Reprinted from \cite{Lin14} {\Ivan under a Creative Commons Attribution 4.0 International License}.\label{Fig_nanodots_MOKE}}
\end{figure}   

\begin{figure}[t]
\includegraphics[width=8.5 cm]{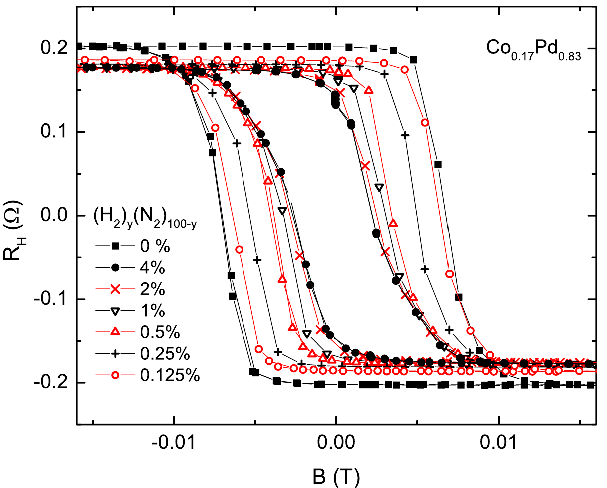}
\caption{Hysteresis loops obtained measuring the extraordinary Hall effect (EHE) resistance a 5-nm-thick Co$_{0.17}$Pd$_{0.83}$ sample in the H$_2$/N$_2$ atmosphere at different hydrogen concentrations: $y$ = 0\%, 0.125\%, 0.25\%, 0.5\%, 1\%, 2\% and 4\%. {\Ivan Reprinted from \cite{Ger17}, with the permission of AIP Publishing.}\label{FigEHE}}
\end{figure}

{\OKB
\section{{\OKB Magneto-electronic} hydrogen sensors based on the magneto-optical Kerr effect (MOKE) {\OKB, anomalous Hall effect and other dc-signal based concepts}\label{sec:MOKE}}
Figure~\ref{MOKE_setup}(a) shows a magneto-optical Kerr effect (MOKE) setup developed to investigate materials intended for the use in hydrogen sensing \cite{Cha19_Taiwan}. Alongside the conventional optical investigations, this setup enables studying a number of important physical effects, including reversible changes in the magnetic properties of the film caused by the absorption of H$_2$ and real-time monitoring of the hydrogen diffusion. 

A reversible nature of the changes in the magnetic properties of the films in an H$_2$ atmosphere was originally observed in Pd/Fe bilayer structures \cite{Lin12}. In that experiment, the thickness of the Pd layer was increased and the MOKE setup gradually adjusted until a point, where a considerable enhancement of the intensity of the MOKE signal was reached after the exposure to 1\,atm of H$_2$. The reversibility of this change was confirmed by cyclic desorption and reabsorption of hydrogen, thereby revealing a practically important sensitivity of the magneto-optical response of a material combining an FM metal (Fe) with a highly hydrogenated NM metal (Pd). Similar results were obtained in the follow-up works \cite{Lin13, Lin13_1}, where a reversible modulation of perpendicular magnetic coercivity due to hydrogenation was investigated in Pd/Fe, Pd/Co and Pd/Ni bilayers \cite{Lin13_1} and Pd/Co/Pd trilayer films \cite{Lin13}.

As the next important step, in \cite{Lin14} the effect of hydrogenation on magnetic coercivity was investigated in perpendicularly magnetised Pd/Co multilayer continuous films and nanodots [Fig.~\ref{Fig_nanodots_MOKE}(a--d)]. Compared with the time-dependent hydrogenation response of the Pd/Co continuous films [Fig.~\ref{Fig_nanodots_MOKE}(e)], the nanodot structures exhibited a much faster hydrogenation response of just a few seconds [Fig.~\ref{Fig_nanodots_MOKE}(f)]. This acceleration was attributed to the high ratio of the exposed surface atoms to the volume atoms in the nanodots compared with the continuous films. Moreover, the measured optical response always exhibited a sharp change within a short period of time and a just minor evolution within a longer period of time. This effect was attributed to an $\alpha$-to-$\beta$ phase transition during the absorption and desorption of H$_2$. In particular, while in the $\alpha$-phase H atoms occupy the interstitial sites of the Pd crystalline lattice, more H atoms enter the crystalline lattice in the $\beta$-phase thus leading to the lattice expansion by approximately 2\%--3\% and concomitant changes in the optical properties of the material.

The hydrogen gas sensing based on extraordinary (or anomalous) Hall effect  is compatible with the existing electronic gas detection technologies and it allows simultaneously measuring two independent parameters affected by H$_2$--resistivity and magnetisation. {\OKB For example, Fig.~\ref{FigEHE} shows the experimental field-dependent hysteresis loops measured using a 5-nm-thick  Co$_{0.17}$Pd$_{0.83}$ sample in the H$_2$/N$_2$ atmosphere at different hydrogen concentrations between 0\% and 4\%. It was found in \cite{Ger17} that thinner film samples would be more suitable for sensing applications because of their higher surface-to-volume ratio and since the absolute value of the measured signal increases when the film thickness is decreased.} This proposed concept has been followed \cite{Tra17, ZheESE_19, Das20, Har20, Har21, Har21_APL, Shi21, Sch21} and its feasibility was further confirmed by demonstrations of detection of low concentration of H$_2$ using thin Pd/Co films as the sensor material \cite{Ger17, Das18, Lia18_ESE, ZheESE_19, Das20}. In particular, it was shown that the EHE sensitivity of optimised samples could exceed 240\% per 104\,ppm at H$_2$ concentrations below 0.5\% in the N$_2$/H$_2$ atmosphere \cite{Ger17}, thus providing a more than two orders of magnitude higher sensitivity compared with the competing resistivity-based sensor architectures.

The results of these fundamental studies laid foundation of several types of advanced hydrogen gas sensor prototypes, including, for example, a device capable of detecting two gases--H$_2$ and CO--using the anomalous Hall Effect \cite{Lia18_ESE}. Such a dual-gas sensor can be used for examining whether methane reformation by steam was completed during H$_2$ production, where CO is a byproduct. While the Hall effect is observed when a magnetic field is applied to a metal through which a current flows, in an FM metal the embedded magnetic moments produce an anomalous Hall effect. Because it depends on both electronic and magnetic properties of the metal, it has been shown that the anomalous Hall Effect serves as a useful read-out mechanism in gas sensors \cite{Lia18_ESE}.

In another paper by the same research group \cite{Hsu18}, [Pd/Fe]$_n$ multilayer thin films were fabricated using e-beam-heated evaporators and their MOKE measurements showed that, under specific experimental conditions, the magnetisation direction of the top Fe layer can undergo a
reversible 90$^{o}$ rotation following an exposure to hydrogen. This result means that the Pd layers that mediate the magnetic coupling between the layers of the structure are sensitive to the hydrogen atmosphere. Subsequently, the investigated [Fe/Pd]$_n$ multilayer system can
operate as a giant magnetoresistance (GMR) H$_2$ gas sensor. More recently, these works have been followed by the demonstrations of H$_2$ gas-mediated magnetic domain formation and domain wall motion in Pd/Co alloy films \cite{Cha18_Taiwan} and observations providing an insights into thermodynamical effects of hydrogen on magnetism \cite{Cha20_Taiwan_alloy}, which should be valuable for the development of hydrogen sensing and storage systems.

Finally, returning to the result from \cite{Cha19_Taiwan} summarised in Figure~\ref{MOKE_setup}(b--d), a real-time monitoring of the hydrogen diffusion was reported using a 50-nm-thick Co$_{25}$Pd$_{75}$ thin film sample covered with a 100-nm-thick ZnO layer. While ZnO is an optically transparent material and therefore using it enables conducting MOKE measurements through a thick ZnO cover layer, the ZnO/CoPd interface is stable at room temperature, that is the covering the CoPd film by ZnO does not cause any significant changes in the magnetic properties of the hydrogen-sensitive alloy. The atomic structure of ZnO is compact so that the H$_2$ molecules cannot penetrate through it, which implies that the ZnO cover layer ensures that there is no direct contact between hydrogen and the measurement equipment. Figure~\ref{MOKE_setup}(c) shows an optical microscope image of the above-discussed sample, the left side of which is not covered by the ZnO layer. The arrow indicates the direction of hydrogen diffusion. Figure~\ref{MOKE_setup}(d) presents the experimental MOKE hysteresis loops taken by illuminating the sample at various areas indicated by the rectangles in Fig.~\ref{MOKE_setup}(c), where one can see that the magnetic hysteresis behaviour depends on the content of hydrogen.
}

\section{{\OKB FMR-based magneto-electronic hydrogen sensors: Main achievements to date}}
\subsection{The origin of {\OKB FMR-based}  hydrogen sensors\label{sec:3.1}}
It follows from the discussions above, before the advent of magneto-optical (Sec.~\ref{sec:MOKE}) and other kinds of gas sensors (Sec.~\ref{sec:2.5_MOKE_Hall}), research efforts had been shaped mostly by investigations of detection mechanisms exploiting changes in electrical resistivity of a gas sensitive material. Indeed, the hydride phase of Pd (Sec.~\ref{sec:2.5}) has an increased electrical resistance and thus various electrical H$_2$ gas sensors utilising this property have been developed \cite{Hub11}. However, typically such sensors have a relatively slow response but attempts to accelerate their response result in increased power consumption \cite{Yan09}. To resolve this problem, it has been suggested that the efficiency of {\OKB sensors relying on electrical resistivity measurements} would be improved, and the probability of irreversible changes to Pd decreased, when the constituent material of the sensor is nanostructured. However, making reliable electrical contacts to individual nanotubes or nanowires is time-consuming and often costly \cite{Yan09}. Most significantly, any electrical sensor is not intrinsically safe because usage of electricity in the presence of H$_2$ can cause fire \cite{Gup12}, and, in general, nano-electronic gas sensors are not an exception from this.

\begin{figure}[t]
\includegraphics[width=13.5 cm]{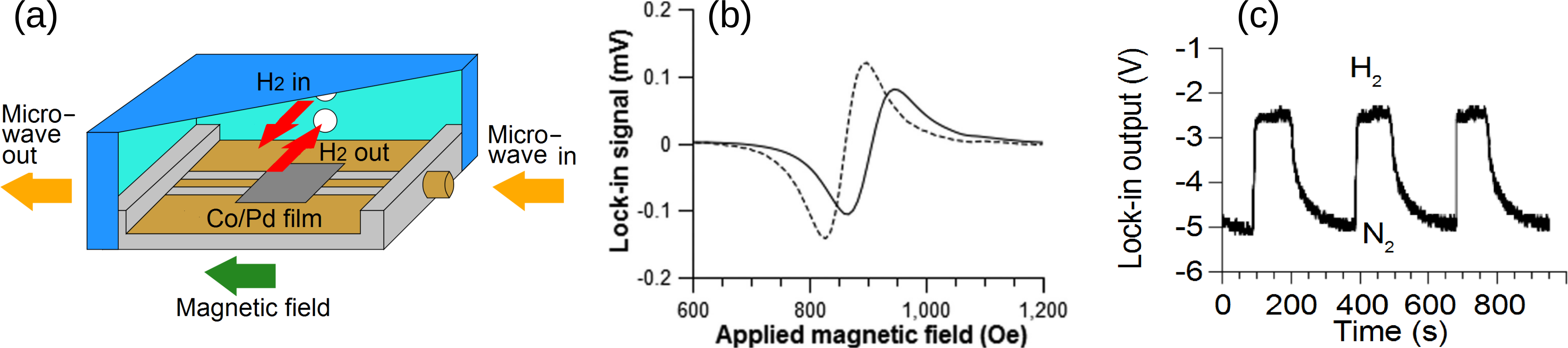}
\caption{\textbf{(a)}~Cross-section of the gas cell showing the co-planar waveguide, where the sample is located, microwave feed ports and gas flow inlets. \textbf{(b)}~Raw FMR spectra of Co[5\,nm]/Pd[10\,nm] at 10\,GHz microwave frequency in N$_2$ atmosphere (solid line) and H$_2$ atmosphere (dashed line). \textbf{(c)}~Change in the microstrip line output voltage under the cycling of N$_2$ and H$_2$ gas through the cell with the Co[5\,nm]/Pd[10\,nm] film operating at the FMR frequency. {\Ivan Reprinted from \cite{Cha13}, with the permission of AIP Publishing.}\label{Fig5}}
\end{figure}

Motivated by the drawbacks of the resistivity-based sensors, a concept using a Pd/FM film and exploiting the physical phenomena of PMA
and FMR was proposed and demonstrated in \cite{Cha13}. The operation of this sensor is based on a change in the strength of PMA upon absorption of H$_2$ by a Pd/FM-metal bilayer thin film and, unlike with any other approach to hydrogen gas sensing, this change is measured using a broadband stripline FMR spectroscopy technique \cite{Mak15}. A customised air-tight cell (schematically shown in Fig.~\ref{Fig5}(a), for technical details see \cite{Thesis_Chris}) was built to enable controlled continuous flow of gas at atmospheric pressure through the experimental setup while performing FMR experiments. The cell contains a co-planar stripline waveguide (CPW), on which the sample under study is located. Coaxial cables feed microwave power from a microwave generator into CPW from one end and carry the transmitted power out through the other end into a microwave receiver. The cell is so fixed between the poles of an electromagnet that the magnetic field is applied along the CPW since this orientation maximises the FMR response \cite{Mak15}. Figure~\ref{Fig5}(b) shows the FMR spectra measured for a Co[5\,nm]/Pd[10\,nm] film under N$_2$ and H$_2$ atmospheres and it demonstrates a readily detectable shift of the resonance peak towards the lower fields upon absorption of H$_2$.

The resulting {\OKB FMR-based} sensor has many advantages. Firstly, fabrication of magnetic multilayers is a well-established and inexpensive process because it relies on the same technological processes that have been used in manufacturing magnetic hard drives for computers. Secondly, a very simple and convenient tool for reading the state of the sensors--the FMR spectroscopy \cite{Mak15}--has been used to reliably detect variations in the SP strength and a reduction of PMA upon exposure of the Pd layer to H$_2$. In the FMR data [Fig.~\ref{Fig5}(b)], a decrease in the SP efficiency was evidenced by the decrease in the width of the resonance line upon filling the cell with H$_2$ but the simultaneous shift in the FMR frequency evidenced a decrease in PMA. Due to the combined action of these two effects, the amplitude of the FMR response, which was set close to the resonance maximum in a N$_2$ atmosphere, decreased after filling the chamber with H$_2$. Significantly, this effect was fully repeatable and reproducible [Fig.~\ref{Fig5}(c)].

{\OKB The authors speculated that} the fabrication of inexpensive on-chip microwave self-oscillators {\OKB was well-established}, for example, in smartphones, WiFi and Bluetooth devices. Hence, this technology {\OKB could} be borrowed to {\OKB implement} FMR detection in the proposed magneto-electronic sensor. Furthermore, in \cite{Cha13}, it was demonstrated experimentally that the sensor state can easily be read out remotely by means of FMR measurements conducted through an optically non-transparent wall of a vessel containing H$_2$. This is in contrast to the magneto-optical techniques (Sec.~\ref{sec:2.6} and Sec.~\ref{sec:MOKE}), where a specially designed and fabricated optically transparent window is required. Significantly, due to a perfect microwave shielding effect in metallic thin films of electromagnetic sub-skin-depth thicknesses \cite{Mak13, Mak14, Mak15}, the microwave electric field applied to the Co side of the bilayer will be practically absent in the space behind the Pd layer (i.e.~inside the vessel containing H$_2$), which eliminates the possibility of arcing and fire.
\begin{figure}[t]
\includegraphics[width=13.5 cm]{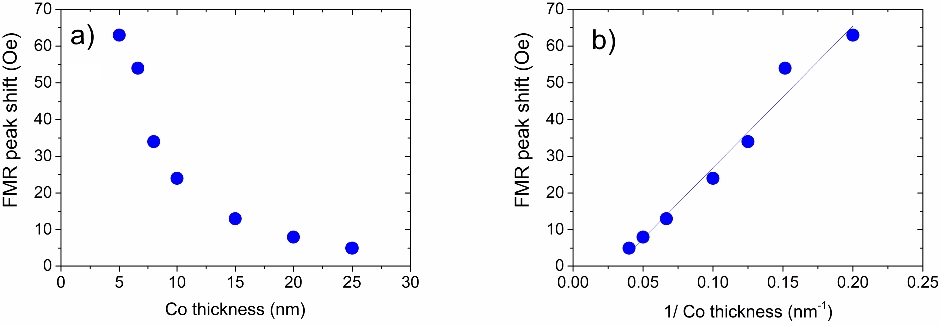}
\caption{\textbf{(a)}~Dependence of the H$_2$-gas induced FMR field shift as a function of the thickness of the Co layer of the samples. The driving microwave frequency is 10\,GHz. The thickness of the Pd layer--10 nm--is the same in all measurements. \textbf{(b)}~The same result as in Panel~(a) but plotted as a function of the inverse thickness of the Co layer, showing a clear linear dependence. Points: experiment, solid line: fit with a straight line. Reprinted with permission from \cite{Thesis_Chris}.\label{Fig6}}
\end{figure}   

Finally, in contrast to many Pd-based optical \cite{Liu11}, electrical \cite{Lee12} and micro-mechanical \cite{Hen12} sensors, the magneto-electronic sensor approach does not require mechanical stretching/shrinking of the Pd layer (see a relevant discussion in Sec.~\ref{sec:3.3}). In general, the strain results in unfavourable physical effects, including a limited lifetime of the sensor due to irreversible layer deformations \cite{Dan02, Lee10} and hysteretic sensitivity \cite{Lee10}. Yet, strain-based sensing requires large Pd thicknesses to overcome the substrate clamping effect \cite{Lee10}, thereby increasing the cost of the sensor since industry-grade Pd is expensive. The modification of PMA does not require micron-scale deformations of the macro-shape of the sensing body. H$_2$ absorption  modifies $d-d$ hybridisation at Pd/Co interface that has been shown to significantly affect the strength of PMA \cite{Miz10, Pal11, Kly20} {\OKB (see Sec.~\ref{sec:2.1} above)}. That is, the sensing effect underpinning the operation of the {\OKB FMR-based} sensor is based on modification of electronic properties of the interface when H$_2$ is absorbed by Pd. This is a considerable advantage because, as shown in \cite{Lee10}, in ultrathin Pd films cyclic hydrogenation does not produce noticeable irreversible deformation of the Pd layer. Therefore, whereas the sensitivity of a typical electronic sensor decreases with decrease in Pd thickness \cite{Lee10}, in the {\OKB FMR-based} magneto-electric sensor the thickness of the Pd layer can be less than 10\,nm \cite{Cha13} (and potentially much less than 10\,nm) due to the interface nature of the PMA and SP processes.

\subsection{The role of the thickness of Pd/Co bilayer films\label{sec:3.2}}
Once the concept of the {\OKB FMR-based}  hydrogen sensors has been introduced, further detailed investigations of the physics underlying its operation were needed. Since the amplitude of the FMR response scales as the volume of the resonating material (Sec.~\ref{sec:2.1}), intuitively it was clear that for sensing applications the Co layer had to be at least a couple of nanometres thick. To provide further insight into the role of the film thickness, in \cite{Commad} the FMR response of Pd-Co bilayer thin films to H$_2$ at atmospheric pressure was experimentally studied with the focus on dependence of the shift of the resonance peak on the Co layer thickness. In the experiments, the thickness of the Pd layer was fixed to be 10\,nm while the thickness of the Co layer was varied from 5 to 25\,nm (Co = 5, 6.6, 8, 10, 15, 20 and 25\,nm). All the samples were sputtered using an in-house sputtering machine \cite{Thesis_Chris}.

FMR field sweeps were taken at fixed microwave frequency of 10\,GHz and resonance field shifts were measured while changing atmosphere between pure N$_2$ and pure H$_2$ gas (Sec.~\ref{sec:3.1}). The larger the FMR peak shift due to hydrogenation, the higher the sensitivity of the sample to H$_2$. A single-layer 5\,nm thick Co film was also prepared to be used as a control sample.

The control sample did not show any FMR peak shift under the exposure to H$_2$. Moreover, FMR peak positions for all the samples were totally reversible while cycling between the 100\% of N$_2$ and 100\% of H$_2$. Overall, the FMR experiments revealed a clear physical behaviour, where the FMR peak shift due to the presence of H$_2$ rapidly decreases with an increase in the thickness of the Co layer (Fig.~\ref{Fig6}). This dependence is consistent with the resonance shift due to a reduction in the strength of {\OKB interface} PMA at the interface of the Co and Pd layers \cite{Cha13, Oka02, Lin13}. 

These results are important for further improvement and optimisation of Pd/Co multilayers. For example, as confirmed by a straight line fit in Fig.~\ref{Fig6}(b), the dependence of the sensitivity to H$_2$ on the thickness of the Co films exhibits a linear behaviour when plotted as a function of the inverse thickness, which speaks strongly in favour of the fact that an interface effect is observed.
\begin{figure}[t]
\includegraphics[width=8.5 cm]{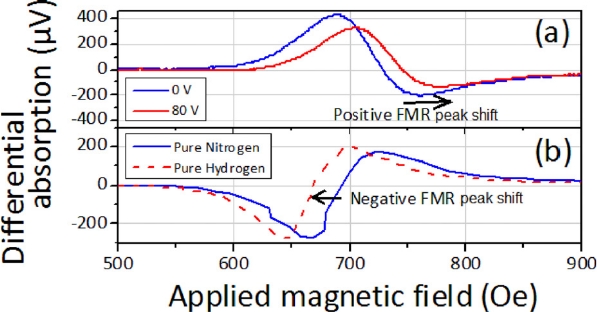}
\caption{Typical examples of the obtained raw FMR traces demonstrating that \textbf{(a)}~application of elastic strain shifts the FMR field upwards, \textbf{(b)}~but exposure of the sample to H$_2$ leads to an FMR field downshift, which is interpreted as a decrease in PMA. In Panel~(a), the FMR-under-stress frequency is 8\,GHz, and the blue (red) curve corresponds to the driving voltage $U = 0$ ($U =80$\,V) applied to the piezoactuator. In Panel~(b), the blue (red) curve correspond to the data taken in the atmosphere of pure N$_2$ (H$_2$) and 10\,GHz frequency. {\Ivan Reprinted from \cite{Lue17}, with the permission of AIP Publishing.}\label{Fig7}}
\end{figure}   

\subsection{Magneto-elastic PMA effect\label{sec:3.3}} 
As demonstrated in Sec.~\ref{sec:2.1}, the PMA is induced at an interface between an FM layer and an NM heavy-metal layer such as Pt or Pd in FM/NM bilayer or multilayer films. The most important physical mechanisms underpinning the interfacial PMA are: (i)~breaking of the crystal symmetry at the interface, (ii)~interface alloying and (iii)~magnetostriction \cite{Hir15}. In (i) and (ii), spin orbit coupling plays an important role and therefore these two mechanisms can be considered to be of electronic nature \cite{Wel94}. However, the third mechanism behind PMA (we call it magneto-elastic PMA \cite{Lue17}) is indirect since the elastic strain at the interface FM/NM mediates the effect of the NM material on magnetisation dynamics in the bulk of the FM layer \cite{Cha88}.
 
It is also well-known that breaking continuity of a thin film through lateral nanopatterning strongly decreases the characteristic time for the transient process of the change in PMA due to exposure to H$_2$ \cite{Lin14, Lue16}. The results of a relevant experiment involving nanopatterned samples \cite{Lue16} suggest that the magneto-elastic contribution may be important, even dominating, since breaking the continuity of the film may allow for easier lateral expansion of the film exposed to H$_2$ atmosphere. On the other hand, bulk Co is known to possess a negative magnetostriction coefficient \cite{Ric69}, and therefore it is legitimate to expect that the resonance field shift due to the magneto-elastic contribution would be positive. However, this contradicts previous observations of a negative {\OKB FMR peak shift} made in in-plane FMR measurements in the presence of H$_2$ \cite{Cha13, Commad, Lue16}.

An attempt to address this discrepancy in experimental data was made in \cite{Lue17}, where FMR measurements of continuous Pd/Co bilayer and Pd/Co/Pd trilayer films grown on flexible Kapton® (Kpt) substrates were conducted with the samples exposed to either H$_2$ gas or elastic stress. In addition, observations of bending of samples due to internal strain induced by the presence of H$_2$ were made. To this end, two extra samples were grown on 21-mm-long and 2-mm-wide Kapton substrates. One of them was a Co[5\,nm]/Pd[10\,nm] bilayer film and the second one a trilayer Pd[20\,nm]/Co[10\,nm]/Pd[20\,nm] film. From the mechanical point of view, such long strips behave effectively as a beam that bends towards the substrate, thereby indicating that the Pd layer stretches. In turn, from the bending radius one can quantify the degree of deformation and then calculate the change in the volume PMA of the Co layer caused by magnetostriction.

In the measurements (Fig.~\ref{Fig7}), exposing the samples to H$_2$ resulted in a downshift of the FMR field. However, FMR measurements conducted in the presence of an externally applied predominantly tensile elastic stress showed an up-shift in the field consistent with negative values of the saturation magnetostriction coefficients for the fabricated samples.

A close inspection of the data obtained in \cite{Lue17} revealed that the magneto-elastic contribution to the H$_2$-induced change in PMA is very small (0.3\,Oe) and that it is of the opposite sign to the electronic contribution (from 20 to 30\,Oe) due to the effect of hydrogen ions on the hybridisation of Co and and Pd orbitals at the interface. A similar mechanism was suggested to explain the change in magnetic properties of CoPd alloys in the presence of H$_2$ \cite{Lin16}. However, in the case of alloys, there is an increase in magnetic moment accompanied by a change in the strength of PMA, which is different physical mechanism. The analysis also showed that the magneto-elastic contribution originates from flexibility of Kapton substrates that bend under the stress induced in the Pd layer by incorporation of H$_2$ into the Pd lattice. In the previous studies (e.g.,~\cite{Cha13, Lue16, Commad}), samples were grown on much more rigid Si substrates with Young’s modulus of more than 130\,GPa, and therefore bending due to the H$_2$-induced stress in the Pd layer was negligible. Accordingly, the contribution of magnetostriction to the FMR peak shift for Pd/Co and Pd/Co/Pd films grown on Si substrates completely vanishes.

{\OKB In the context of the discussion above, it is interesting to note that in H$_2$ sensors exploiting plasmon resonances in metal nanostructures (Sec.~\ref{sec:2.6}), an H$_2$-induced spectral shift consists of two competing contribution: a small down-shift in the resonance optical wavelength caused by changes in the optical dielectric function of Pd and a much stronger up-shift caused by expansion of the Pd lattice \cite{Tit12}. Based on these results, initially the authors of this study suspected that the operation of plasmonic sensors could be affected by strain effects (they explained a discrepancy between their experimental and numerical results by the fact that strain effects were not considered in their model). However, in their followup work \cite{Str13}, the same authors demonstrated that their improved sensor architecture, which consists of a PdNi alloy film, calcium fluoride buffer and Pt capping layer, exhibits good temporal stability while still providing a strong signal in response to exposure to H$_2$ gas. In particular, it was shown that while the use of a calcium fluoride buffer reduces the stress of the PdNi alloy film, the Pt capping layer helps avoiding the adverse effect of surface poisoning. While these results might not be of direct relevance to FMR-based hydrogen sensors, the methods reported in these works can be adopted in the MOKE hydrogen sensors because they share the same photonics-based signal registration concept underpinning the operation of plasmonic sensors \cite{Mak15_review, Mak16_1}.} 

To conclude this section, we mention that a theoretical model of strain in the Co layer in the presence of H$_2$ was proposed in \cite{Lue17}. Normally, the analysis of stress requires conducting rigorous numerical simulations using either a finite elements or finite-difference method \cite{Mak_SciReps}. Significantly, outcomes of any modelling depend on input mechanical material parameters that, strictly speaking, can be very different in case of thick and thin films and nanostructures (see \cite{Mak_SciReps} and references therein). However, numerical modelling may be avoided in the case of bilayered materials because analytical progress can be made using Stoney formula known from the field of semiconductor structures, where the stress in a thin film results in the buckling of the wafers and the radius of the curvature of the stressed structure becomes related to its stress tensor \cite{Suo99}. It was assumed that the strain in the buried Co layer is the same as in the capping Pd layer, which is a warranted simplification given that Young’s modulus of thin-film Co (approximately 200\,GPa \cite{Kal19}) is similar to that of thin-film Pd (approximately 120\,GPa \cite{Jen05}) and {\Ivan two} orders of magnitude higher than that of Kapton substrates (4\,GPa).

FM films on Kapton develop a strain during their fabrication process. This natural cylindrical bending was visible with the naked eye in \cite{Lue17} and its radius could easily be measured before the experimental gas cell was filled with H$_2$. Hence, the actual strain due to H$_2$ absorption was evaluated by subtracting the strain due to the natural bending from the strain measured in the presence of H$_2$. Using this procedure, it was established that the strip was {\OKB originally} bent in the opposite direction with respect to the bending direction due to H$_2$ absorption, which means that originally the metallic layers were compressed. In turn, absorption of H$_2$ reduced the negative strain and increased the bending radius.
\begin{figure}[t]
\includegraphics[width=12.5 cm]{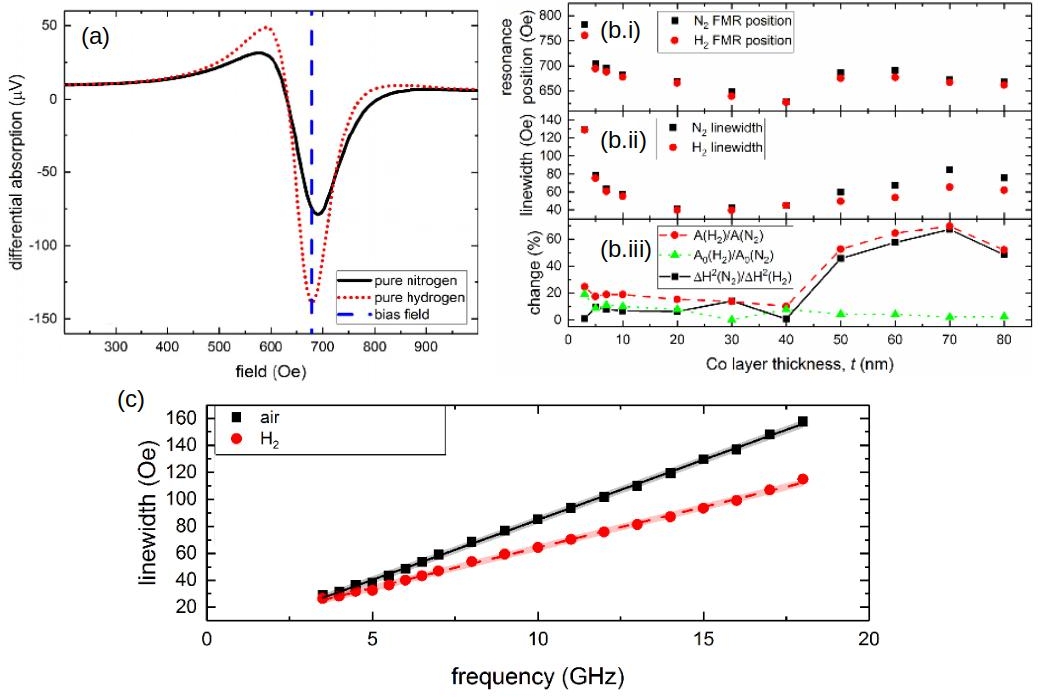}
\caption{\textbf{(a)}~Differential FMR traces for sample with the Co layer thickness $t=70$\,nm in N$_2$ (black solid curve) and H$_2$ (red dotted curve) environments. The blue dashed line highlights the point of the biggest change (69.5\%) in the FMR response. Panels~(b.i) and (b.ii) show the resonance fields and linewidth of the FMR peak as a function of the Co layer thickness $t$ at $f=10$\,GHz in N$_2$ (black squares) and H$_2$ (red dots) environments. Panel~(b.iii) shows the changes of the linewidth and FMR amplitude given in percentage. \textbf{(c)}~Linewidth $\Delta H$ of the FMR peak as a function of the microwave frequency $f$. The black squares denote $\Delta H$ in air atmosphere and the red dots denote $\Delta H$ in H$_2$ environment {\OKB for the 70-nm-thick sample}. The straight lines are the respective linear fits.  {\Ivan \copyright~2018 IEEE. Reprinted, with permission, from \cite{Sch18}.}
\label{Fig8}}
\end{figure}   

\subsection{Effect of H$_2$ gas on the FMR response of Pd/Co structures\label{sec:3.4}}
Thus, to summarise the key results discussed in the previous sections, it has been demonstrated that H$_2$ gas can be reliably detected using a Pd/Co bilayer system driven by means of FMR and that the operating principle of the resulting sensing device is based on changes in PMA existing at the interface between the Co and Pd layers, where the strength of PMA decreases when the Pd layer is exposed to H$_2$. It was also established that the physical processes underlying the change in the strength of PMA are of electronic nature and not due to magnetostriction effects, which overall enables observing measurable shifts in the resonance peak of the FMR spectrum. Furthermore, the effect of the interface PMA on the FMR response of a Pd/Co system is maximised when the thickness of the Co layer is small. For example, a typical thickness of the Co layer used in many experiments discussed thus far is $t = 5$\,nm, which is, in general, chosen as a trade-off between an increase in the interface PMA effect caused by a decrease in the Co layer thickness $t$ and a decrease in the amplitude of the resonance peak (i.e.~a decrease in the sensitivity of measuring electronics to the presence of absorption of microwave power by Co in the conditions of FMR) that is caused by a decrease in the volume of the FM material.
 
However, Pd/Co structures exhibit other intriguing effects that are relevant to hydrogen sensing but their physical nature has not been fully understood yet. For instance, in the works \cite{Cha13, Commad, Lue16, Lue16_1, Lue17_1}, where the foundations of {\OKB FMR-based} sensors were laid, an increase in the amplitude of the resonance peak was observed alongside a shift in the position of the FMR peak caused by the exposure to H$_2$. The increase in the amplitude was accompanied by a decrease in the linewidth of the FMR peak.

A systematic study of this effect was conducted in \cite{Sch18}, where Pd/Co structures with a Co layer of varied thickness $t$ ($t$~=~3, 5, 7, 10, 20, 30, 40, 50, 60, 70 and 80\,nm) sandwiched between a 10-nm-thick capping Pd layer and a 10-nm-thick technologically important tantalum (Ta) seed layer \cite{Mak13} were investigated. In the presence of H$_2$, it was established that for the samples with the Co layer thicker than 50\,nm the shift of the FMR peak quickly decreases. Conversely, the hydrogen-induced decrease in the FMR peak linewidth remains for large Co thicknesses, and fully correlates with an increase in the FMR peak amplitude that was also observed for the whole thickness range for the studied films. For example, Fig.~\ref{Fig8}(a) shows the differential FMR traces for a $t = 70$\,nm sample in N$_2$ and H$_2$ gas atmospheres, where there is a significant difference in the amplitude for H$_2$ and N$_2$ environments: the peak amplitude increases by 69.5\% in the presence of H$_2$ and there is virtually no H$_2$-induced FMR peak shift. Subsequently, it is plausible to assume that different physical phenomena underlie a peak shift and a change in the amplitude.

Thus, while an increase in the resonance shift with the decrease in $t$ was observed in \cite{Commad} for thin ($t < 25$\,nm) Co layers and interpreted as a decrease of the contribution of the interface PMA that scales as $1/t$, in \cite{Sch18} the samples with the Co layer thickness $t > 40$\,nm did not show any $1/t$ dependence [Fig.~\ref{Fig8}(b.i)], thereby impeding the analysis of the sensor performance in terms of contributions of PMA. At the same time, it was observed that the FMR linewidth $\Delta H$ [Fig.~\ref{Fig8}(b.ii)] decreases as $1/t$ in the Co layer thickness range from 3 to 20\,nm (for pristine samples) and that then it linearly increases for thicker films with $t > 20$\,nm. Figure~\ref{Fig8}(b.iii) provides a quantified measure of these changes, where the linewidth $\Delta H$, amplitude $A$ and parameter $A_0$, which represents the "excitation amplitude" that combines the strength of coupling of the magnetisation vector precession to the driving microwave magnetic field and the amplitude of the driving field, are plotted as a function of $t$ at $f = 10$\,GHz for N$_2$ and H$_2$ environments.

Furthermore, Fig.~\ref{Fig8}(c) shows the linewidth $\Delta H$ measured for the sample with a $t = 70$\,nm Co film as a function of the microwave frequency $f$ in air and in H$_2$ atmosphere, where one can clearly observe a change in the slope of the curve in the presence of H$_2$ compared with that in air. Assuming that $\Delta H$ is the half-linewidth of a Lorentzian-shape resonance peak, it can be shown that $\Delta H = \Delta H_{0}/2 + \alpha_G f/|\gamma|$ (see \cite{Mak15}), where $\Delta H_0$ is a measure of inhomogeneous broadening, $\alpha_G$ {\OKB is} Gilbert damping constant and $|\gamma|$ is the absolute value of the gyromagnetic ratio. Because the  
{\OKB positions} of the resonance peak in N$_2$ and H$_2$ environments are only slightly different [Fig.~\ref{Fig8}(b.i)], it is plausible to assume that parameter $\gamma$ is also unchanged. Hence, the {\OKB difference in the slopes of the straight lines} in Fig.~\ref{Fig8}(c) must be mostly due to {\OKB a difference in values of $\alpha_G$}.

It is well-known that $\alpha_G$ originates from several different physical processes \cite{Mak15}. For instance, it is material specific and therefore it describes the linewidth broadening due to magnon-magnon scattering and loss of energy to the crystal lattice \cite{Mak15}. However, these physical mechanisms alone cannot give satisfactory explanation of the dependence on the Co film thickness observed in Fig.~\ref{Fig8}(b) because H$_2$ does not penetrate into the bulk of the Co layer and therefore all changes to $\alpha_G$ induced by H$_2$ take place in the Pd layer and at the interface between the Pd and Co layers.

Indeed, when Pd is in contact with an FM material, a magnetic moment is induced on Pd atoms that are in a direct contact with atoms of the FM material. Known as the magnetic proximity effect (MPE) \cite{Kim16}, this process can lead to an increase in $\Delta H$ \cite{Con16}. Pd/Co structures also exhibit the SP effect that results in an increase in $\Delta H$ \cite{And11_1, Wat18}, because, due to its small spin diffusion length, Pd serves as a good spin sink, when the thickness of its layer is of order of 10\,nm \cite{For05, Sha12, Cam16}. Therefore, the presence of SP processes can explain the $1/t$ dependence of the linewidth for $t \le 20$\,nm in Fig.~\ref{Fig8}(b). Moreover, in the experiments in \cite{Roj12} the contributions of the MPE and SP effects to $\alpha$ in Co/Pt structures were notably equal when the Pt layer was thicker than 2\,nm. Although Pt exhibits stronger spin-orbit interaction with stronger SP processes, the relative contribution of the MPE and SP in Pd should be qualitatively similar. Since MPE is an interface effect, it may contribute to the $1/t$-dependence observed for small values of $t$. However, it cannot explain linear increase of $\Delta H$ in the case of $t > 20$\,nm either. 

{\OKB We note that in a recent study \cite{Wei21} the impact of H$_2$ gas on the FMR linewidth of the Pd/Y/Ni$_{80}$Fe$_{20}$­ trilayer systems has been found to be different. In particular, it has been observed that the absorption of H$_2$ by Y resulted in a decrease in the frequency-independent contribution to the FMR linewidth $\Delta H_0$ and it has not affected the Gilbert damping constant $\alpha_G$. Parameter $\Delta H_0$ is known to be affected by the structural and material quality of the film surfaces and interfaces and, therefore, it has been speculated that the decrease in $\Delta H_0$ could be due to an interface clearing effect \cite{End05}. The effect results in an improvement of the quality of the Y--to--Ni$_{80}$Fe$_{20}$ interface under the influence of H$_2$ gas. Significantly, the observed change in the resonance linewidth was fully reversible--the linewidth returned to the same larger value on evacuation of the gas from the sample environment. Remarkably, this implies that the interface clearing was fully reversible.}
     
The interface clearing effect was first reported in \cite{Kam02, End04, End05}. Those works investigated the effects of hydrogenation on the structure, transport and magnetic properties of FM/NM (FM = Fe, Co and NM = La, Y, Gd) structures with layer thicknesses comparable with those from \cite{Sch18}. It was demonstrated (i)~that the behaviour of the NM layers can transition from metal to semiconductor, (ii)~that the NM layers expand and (iii)~that the saturation magnetisation of the FM layer increases. Therefore, it was concluded that hydrogenation represents a practicable method for changing the interface state and producing a semiconductor layer inside a multilayer structure, thereby opening opportunities for producing new functional magnetic multilayers. Although these conclusions were drawn for the structures using rare-earth metals as an NM material, similar behaviour is expected in the case of a NM layer made of Pd because {\OKB of a similarity of its} response to hydrogenation. Significantly for this present discussion, in \cite{Kam02, End04, End05} it was suggested that the FM/NM interface becomes more abrupt (i.e.~"clear") as a result of hydrogenation. In other words, effectively the adsorption of hydrogen improves the properties of the interface and, as a result, hydrogen does not enter the FM layer (i.e.~the Co layer of the magneto-electronic sensor). This process is reversible and, since the quality of the FM material becomes effectively improved, a narrower FMR linewidth and a higher FMR amplitude are observed. Although the NM layer expands as a result of hydrogenation, and thus it would be plausible to expect that this would influence the structure of the adjacent FM layers through the FM-NM interface, X-ray diffraction measurements did not reveal any FM layer deformation \cite{Kam02}.

Returning to the results from \cite{Sch18}, despite the uncertainty surrounding the physical origin of the effect of H$_2$ on the amplitude of the response of the Pd/Co interfaces, this work demonstrated that this effect can be used to measure H$_2$ concentrations from 0.05\% to 70\% with a higher sensitivity than magneto-electronic sensors using a shift in the {\OKB FMR peak position}. 
\begin{figure}[t]
\includegraphics[width=12.5 cm]{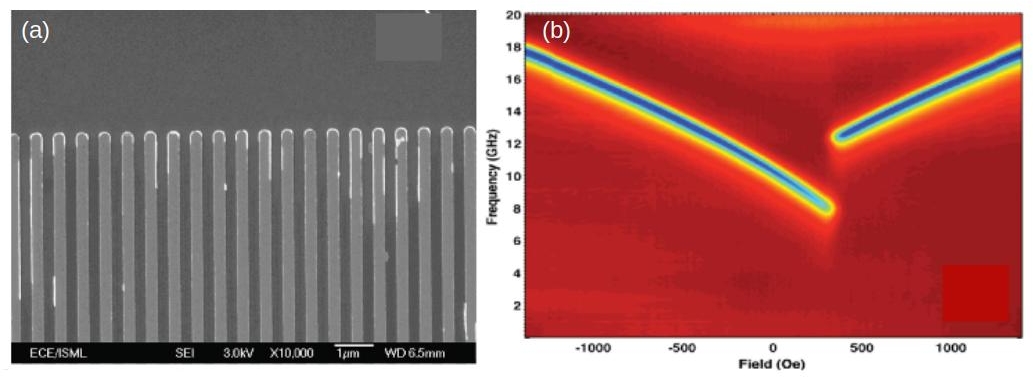}
\caption{\textbf{(a)}~Scanning electron microscopy image of the Pd/Co/Pd nanowires. \textbf{(b)} False colour map of the FMR response of the as-deposited nanowire array sample (the plot shows the FMR amplitude as a function of the applied field and frequency). Reprinted with permission from \cite{Lue16}.
\label{Fig9}}
\end{figure}   

\subsection{Nanopatterned {\OKB FMR-based} sensors\label{sec:3.5}} 
It is well-established that nanopatterning of the active Pd element can increase the sensitivity of hydrogen gas sensors that exploit changes in the resistivity of Pd due to hydrogenation (see, e.g.,~\cite{Fav01, Yan09, Lim15}). However, nanopatterning does not change the principal physical mechanisms underlying the detection of H$_2$ in this kind of sensors and therefore it does not eliminate the possibility of arcing and fire due to voltage applied to sensor.
   
In the work \cite{Lue16}, it has been demonstrated that {\OKB FMR-based  sensors} using nanopatterned Pd/Co films also exhibit considerable advantages over sensors based on continuous Pd/Co films. In particular, in \cite{Lue16} it has been shown that nanopatterning of Pd/Co films results in a higher sensitivity to H$_2$ and a much faster hydrogen {\OKB absorption and} desorption rates. Significantly, nanopatterning also allows avoiding the need for applying an external biasing magnetic field, which may be important for practical sensor implementation since this considerably {\OKB reduces the fabrication costs}. A wide range of H$_2$ concentrations from 0.1\% to {\OKB 60}\% were detected using the proposed nanopatterned Pd/Co, with a special focus on the 4\% threshold of hydrogen flammability in air. {\OKB Note that 60\% of H$_2$ gas was not a natural limit for this medium. Potentially, hydrogen-gas concentrations above 60\% could be resolved if the sensor was retuned as suggested later in \cite{Lue17_1}}.

As discussed in Sec.~\ref{sec:2.2}, an {\OKB FMR-based} sensor relies on the phenomenon of FMR, and therefore its operation would be impossible without application of a static magnetic field. Indeed, the FMR frequency for a continuous FM film at {\OKB a} zero external magnetic field is zero \cite{Mak15}. Consequently, only by applying an external static magnetic field one can obtain a non-vanishing FMR frequency and hence enable a finite-frequency FMR response of the sensor.

However, already in the pioneering work \cite{Cha13} it was suggested that nanopatterning of the active element of the {\OKB FMR-based} magneto-electronic sensor might eliminate the need for an external static magnetic field. To confirm that theoretical prediction, in the experiments in \cite{Lue16} a continuous Pd/Co film was replaced by a periodic array of macroscopically long trilayer Pd[10\,nm]/Co[20\,nm]/Pd[10\,nm] structures [Fig.~\ref{Fig9}(a)] of nanopatterned strips ("nanowires") with a sub-micrometer cross-section. Having a width of 300\,nm and a pitch of 600\,nm, these nanowires were fabricated using deep-ultraviolet lithography followed by electron beam layer deposition and, subsequently, lift-off \cite{Sin04}. A reference continuous film with the same trilayer composition was also prepared. All structures were investigated using the standard in-plane FMR spectroscopy setup \cite{Mak15}.
\begin{figure}[t]
\includegraphics[width=12.5 cm]{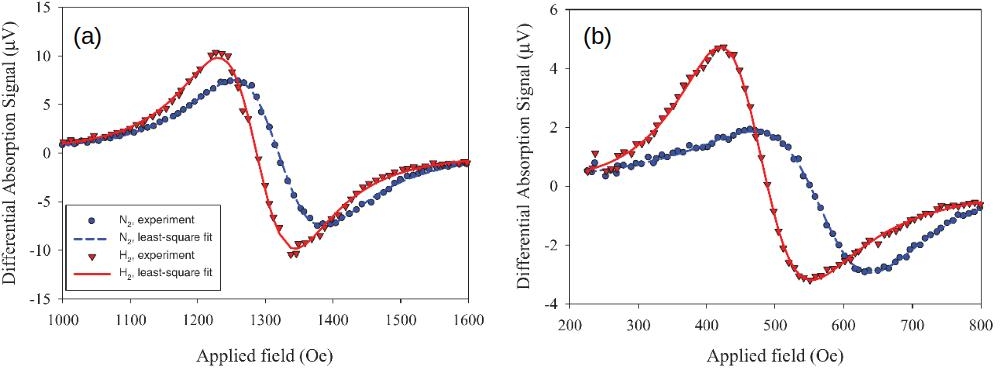}
\caption{\textbf{(a)}~FMR response at large applied fields for the continuous film and \textbf{(b)}~for the nanowires at 13\,GHz microwave frequency. The circular dots and triangles denote the experimental data obtained in H$_2$ and N$_2$ environments, respectively. The solid and dashed curves are the respective fits with a complex Lorentzian. Reprinted from \cite{Lue16}.
\label{Fig10}}
\end{figure}   

Figure~\ref{Fig9}(b) shows the FMR response of an as-deposited structure, where the applied static magnetic field and the microwave frequency were so scanned over a broad range of values that the entire width of the major hysteresis loop for the sample was characterised. The shape of the hysteresis loop for the nanowires is close to a square one because the response of the nanowires switches sharply between the single-domain states, where the magnetisation vector points along the nanowire axis \cite{Din11}. Therefore, a sharp jump in the FMR frequency takes place when the magnetisation vector for the nanowires switches from the negative direction to the positive one [Fig.~\ref{Fig9}(b)]. Importantly, at remanence ($H = 0$), the ground state of the static magnetisation vector is the uniform magnetisation with the magnetisation vector orientated along the long axis of each nanowire. This is due to a large shape anisotropy energy originating from the geometric confinement in the direction of the nanowire width. In addition to being responsible for a large remanent magnetisation, the shape anisotropy provides a contribution to the energy of magnetisation-vector precession, which shifts the FMR frequency up from zero. For instance, in \cite{Lue16} this resulted in an FMR frequency of approximately 10\,GHz for the zero applied field. This large {\OKB FMR} frequency is due to a large dynamic demagnetising (dipolar) field that the precessing magnetisation vector creates in this strongly confined geometry \cite{Gus02, Mak15, Din11}.

Typical differential FMR responses of the continuous film sample and the nanopatterned sample are shown in Figs.~\ref{Fig10}(a) and (b), respectively, where the microwave frequency was set to be 13\,GHz and the measurements were conducted in pure N$_2$ and H$_2$ environments. As with previous similar measurements of continuous film-based sensors, results of which have been discussed in the preceding sections, a significant shift of the FMR peak to the lower fields as well as narrowing of the resonance line are observed as a result of hydrogenation for both types of samples. However, significantly, the nanowire array exhibits a superior H$_2$ gas sensing performance both in terms of the absolute values of the extracted FMR parameter variations and percentage change with respect to the as-deposited state of the sample. Indeed, one can see that the shift in the resonance peak position is more than three times higher in the nanowire sample, resulting in a readily detectable shift that is equivalent to about 75\% of the resonance linewidth.

\begin{figure}[t]
\includegraphics[width=7.5 cm]{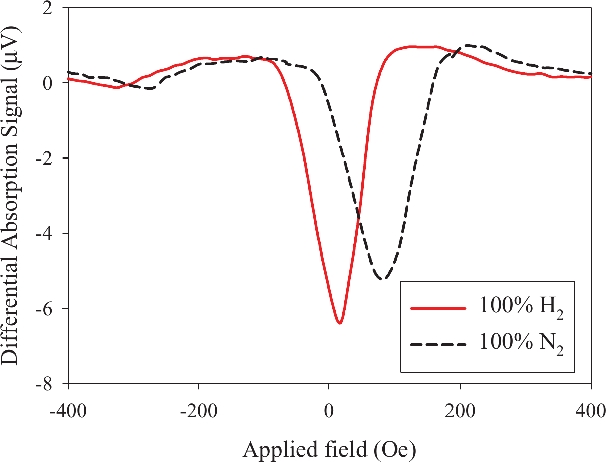}
\caption{FMR responses of a nanowrire Pd/Co/Pd structure at the microwave frequency 10.2\,GHz for a pure N$_2$ gas environment (dashed curve) and pure H$_2$ gas environment (solid curve). Reprinted with permission from \cite{Lue16}.
\label{Fig11}}
\end{figure}

Figure~\ref{Fig11} shows the field-resolved FMR traces scanned across zero applied field at 10.2\,GHz, where one can see that for $H = 0$ the FMR amplitude is finite in the presence of H$_2$, but in the N$_2$ environment it is nearly zero. Subsequently, by measuring the amplitude of the FMR response at $H = 0$ one can reliably detect the presence of H$_2$ gas in the atmosphere. Furthermore, the variation in the response amplitude carries information about the concentration of H$_2$ in the environment.
 
\begin{figure}[t]
\includegraphics[width=13.5 cm]{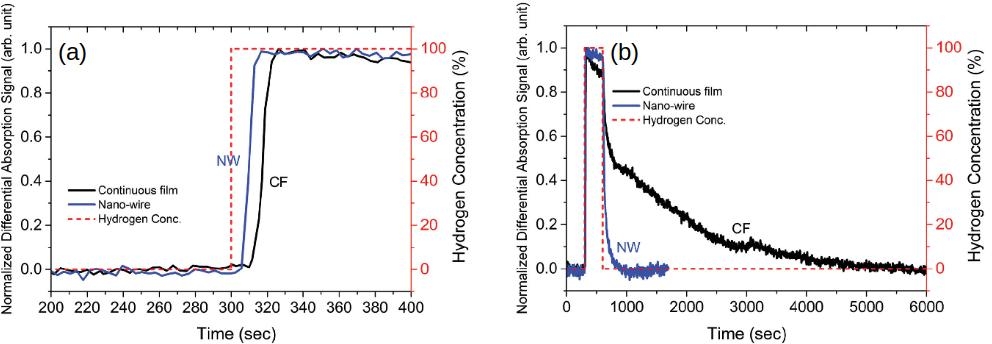}
\caption{\textbf{(a)}~Hydrogen absorption trace measured over the first 100\,s after letting H$_2$ gas into the test chamber. \textbf{(b)}~The same signal as in Panel~(a) but plotted over an extended period of time and showing both the absorption and desorption traces. The labels "NW" and "CF" denote the traces for the nanowires and the reference continuous film, respectively. The red dashed curve indicated the sharp change in the gas atmosphere inside the chamber with the samples. Reprinted with permission from \cite{Lue16}.
\label{Fig12}}
\end{figure}

Let us now compare and discuss the rates of desorption and absorption of H$_2$ by the nanopatterned and continuous film-based sensors. To enable such a comparison, the static magnetic field was fixed to be 510\,Oe for the nanowire sample and 1280\,Oe for the continuous film sample, the microwave frequency was set to be 13\,GHz and the amplitude of the FMR signal was monitored as a function of time (Fig.~\ref{Fig12}). The test gas chamber was first flushed with N$_2$ for five minutes and then the atmosphere inside it was changed to pure H$_2$ gas. As shown in Fig.~\ref{Fig12}(a), in an FMR experiment, this change in the gas environment (the dashed red curve) is seen as a sharp increase in the differential FMR absorption amplitude of both the nanowires sample (the curve labelled as "NW") and the continuous film sample (the curve labelled as "CF") at about 300\,s. Then, after five minutes of hydrogenation the atmosphere within the chamber was changed back to N$_2$ and this resulted in a sharp drop in the FMR signal of the nanowire but a much more gradual decrease in the FMR signal for the continuous film [Fig.~\ref{Fig12}(b)].
 
Thus, the results in Fig.~\ref{Fig12} demonstrate that both absorption and desorption rates for the nano-patterned material are much smaller than for the continuous film. In particular, it was established that for the nanowires at 60\% H$_2$ the absorption time is 22\,s while desorption occurs over 60\,s. This is more than 30 times smaller than for the continuous film and this observation is also in good agreement with a previously observed acceleration of the magneto-optical response of nano-patterned materials with a similar material composition \cite{Lin14}. Significantly, in the case of the {\OKB FMR-based} sensors this acceleration was achieved without any optimisation of the material and geometry of the nanopatterned structure. Indeed, the structure shown in Fig.~\ref{Fig9}(a) was fabricated according to the specification developed for other than sensing purposes and fabricated using specific available equipment. Therefore, it is plausible to assume that the response time can be reduced to the industry requirement response time of one second by optimising the geometry of the nanopattern. Indeed, this idea was recently confirmed by measuring the FMR response of Pd/Fe$_3$O$_4$ core-shell nanoparticles \cite{Sha_unpubl}. {\Ivan The time of the response of the FMR peak for the nanopowder to insertion of just 3\% of hydrogen gas into its environment amounted to just 2\,s.}        

\subsection{Alloy-based magneto-electronic sensors\label{sec:3.6}}
As discussed in Sec.~\ref{sec:2.6}, not only multilayers but also alloys of Pd with Co, Fe and Ni have been intensively investigated for applications in hydrogen sensing \cite{Lin12, Lee12, Lin13_1, Lin16, Win15, Ger17, Hsu17, Lia17_Taiwan, Aka19, Wan18_Taiwan, Cha18_Taiwan, Cha20_Taiwan_alloy, Lue19}. In the cited and other papers by the same research groups, either Anomalous Hall Effect \cite{Ger17}, magneto-optical Kerr effect (e.g.,~\cite{Lin12, Lin13, Lin13_1, Lin14}), also see Sec~\ref{sec:MOKE}) {\OKB, or} FMR effect \cite{Lue19} have been employed to detect H$_2$ gas in the environment of the alloyed films. In particular, in \cite{Lue19} the FMR response of Co$_x$Pd$_{1-x}$ alloy-based sensors with a varying material composition parameter $x$ ($x$ = 0.65, 0.39, 0.24 and 0.14) was investigated and it was found that there exist significant differences in the FMR response of these samples to hydrogenation compared with multilayer based sensors.

\begin{figure}[t]
\includegraphics[width=7.5 cm]{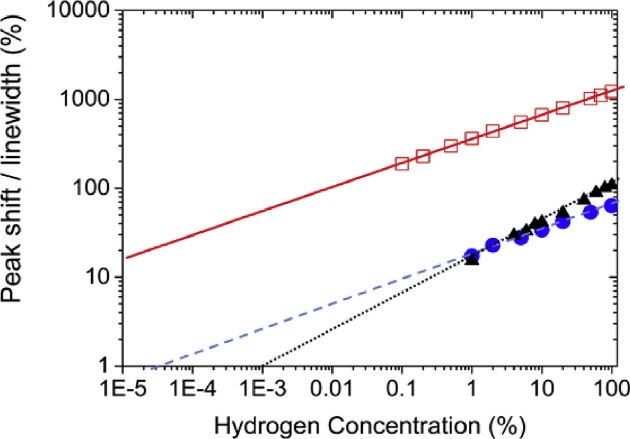}
\caption{Response to H$_2$ gas of the Co$_x$Pd$_{1-x}$ alloy-based sensors with material composition parameter $x = 0.24$ (squares) and $x = 0.39$ (circles) compared with the response of a high-performing Pd/Co bilayer film-based sensor (triangles). All data are plotted on a logarithmic scale and the straight lines are the linear fits to the experimental points. {\Ivan Reprinted from \cite{Lue19}, with permission from Elsevier}.
\label{Fig13}}
\end{figure}

Let us first recall that two kinds of hydrogen sensors are required for applications in hydrogen-gas fuel cells and relevant technologies--safety sensors and concentration meters--where the safety sensors must detect the presence of trace (ppm) amounts of H$_2$ but the concentration meters have to be able to resolve difference in H$_2$ concentrations from {\OKB small values} to 100\%. Alloy-based {\OKB FMR-based} sensors {\OKB were shown to be able to} measure H$_2$ concentration in a very broad range from 0.05\% to 100\%. Moreover, as one can see from Fig.~\ref{Fig13}, the concentration resolution of alloy-based sensors {\OKB employing FMR} is non-vanishing up to 100\% concentration and it scales approximately as $1/C$, where parameter $C$ denotes the concentration of H$_2$, also exhibiting no signature of saturation near this mark. This is in contrast to the entire group of Pd-based solid-state sensors, where many devices saturate at relatively low H$_2$ concentrations (see, e.g.,~\cite{Sun16}). Subsequently, it is plausible to recommend employing alloy-based magneto-electronic sensors as H$_2$ concentration meters operating inside a fuel cell \cite{Hub14}.

{\OKB It is noteworthy that in contrast to our previous discussion of FMR-based studies of the sensors relying on interface PMA processes, in the alloy-based sensors the hydrogen detection principle is based on modification of the strength of the effective field of the bulk PMA. The advantage of using the bulk PMA is that its contribution to the FMR frequency does not depend on the film thickness. On the other hand, for any sample, the amplitude of the FMR peak scales as the volume of the resonating material. In the case of the stripline FMR of magnetic films, this scaling implies a linear increase in the FMR peak height with the film thickness. The higher the peak, the simpler electronic circuitry needed to detect the presence of FMR absorption by the sample. This potentially simplifies the design of a future fully integrated sensor.  

In addition, measurements of the alloyed films fabricated in \cite{Lue19} demonstrated that the values of the FMR linewidth can be in a broad range consistently with the plausible assumption that the resonance line for Co would broaden when we dilute this FM material with non-magnetic Pd. FMR measurements of a film with a broad linewidth of approximately 700\,Oe showed no noticeable increase in the height of the FMR peak for the alloy with respect to the standard Pd[10\,nm]/Co[5\,nm] bilayer film. This is because the peak height also scales as the inverse linewidth thereby compensating the effect of the larger magnetic-layer thickness. However, it appears that the alloy sample that exhibited the larger linewidth is particularly suitable for amplitude-based sensing: for the whole range of investigated H$_2$ concentrations (1\% to 100\%) the FMR peak shift never exceeded the FMR linewidth. This is advantageous from the practical point of view because, in contrast to the conventional bilayer films \cite{Lue17}, there is no need to re-adjust the sensor several times to cover the entire investigated range of H$_2$ concentrations. Significantly, if the FMR linewidth for the standard layered sample was artificially broadened to enable the same adjustment-free operation, the height of the FMR peak would become too small, thereby forbidding the use of simple electronic circuitry because any reliable peak detection in the presence of noise would require the deployment of a complex signal detection system. In other words, using thicker alloyed samples allows one to increase the FMR linewidth without compromising the ability to detect the FMR response itself, thus enabling re-adjustment-free measurement of a broad range of H$_2$ concentrations.
  
On the other hand, the same paper \cite{Lue19} also reports the results of a study of a CoPd alloy film with a significantly narrower FMR linewidth of approximately 250\,Oe. It was found that this sample exhibits an H$_2$-induced FMR peak shift that exceeds its linewidth by a factor of 10 at 100\% H$_2$ concentration. This feature makes such sample particularly suitable for frequency-based sensing, where the H$_2$ concentration becomes effectively encoded in the FMR frequency shift caused by the hydrogenation. Importantly, because of the small peak linewidth, the reported peak height was about 7 times larger than for the sample with the larger linewidth discussed above.}

\subsection{Nickel-cobalt-palladium alloy magneto-electronic sensors\label{sec:3.7}}
Nickel is an important additive material used in hydrogen sensors bases on changes in the electric resistivity or optical response of the active layer. Among other technological roles such as corrosion protection and increased mechanical strength \cite{Cha94, Hug95, Str13}, the addition of Ni helps (i)~accelerate the absorption and desorption processes of H$_2$ gas and (ii)~avoid electric-resistivity response saturation above approximately 20\% of H$_2$ in the environment, which occurs due to a transition from the $\alpha$ phase of Pd to the "hydride" $\beta$ phase and concomitant irreversible behaviour observed both in thin films \cite{Hug92, Hug95, Cha94} and nanoparticle systems \cite{Str13}. It was also demonstrated that the addition of Ni results in a contraction of the Pd lattice \cite{Lew67}. By means of that effect, for bulk samples of Pd-Ni alloys, the solubility of hydrogen decreases and the phase transition to the $\beta$ phase is suppressed. This implies that in 8-15\% Ni alloys significantly higher H$_2$ gas concentrations are required to trigger saturation \cite{Hug92}.

Therefore, it is plausible to assume that the use of Ni in {\OKB Co-Pd} structures of magneto-electronic sensors should provide similar benefits for the detection of H$_2$, also revealing novel fundamental physical effects underlying the operation of such sensors. This idea was verified in the work \cite{Sch21}, where Ni-Co-Pd alloy films of various compositions were investigated using the FMR spectroscopy in air and in H$_2$ environments. It is worth noting here that Ni is a ferromagnetic transition metal, similarly to Co and Fe. It is characterised by the smallest saturation magnetisation and the largest FMR linewidth among these three transition metals. A behaviour similar to that of Co-Pd alloy thin films was observed at high concentrations of H$_2$. However, while the response of Co-Pd thin films steadily increased as the H$_2$ concentration was reduced, for lower H$_2$ concentrations--in the range from 0\% to 5\%--the majority of the investigated Ni-Co-Pd samples demonstrated a sudden decrease in the sensitivity. Moreover, when compared with the response of Co-Pd films, it was found that the Ni-Co-Pd films are characterised by a lower normalised FMR peak shift at a higher Pd content and a larger peak shift at a lower Pd content. This behaviour was attributed to the following differences in fundamental processes occurring in the constituent materials of the films.

\begin{figure}[t]
\includegraphics[width=12.5 cm]{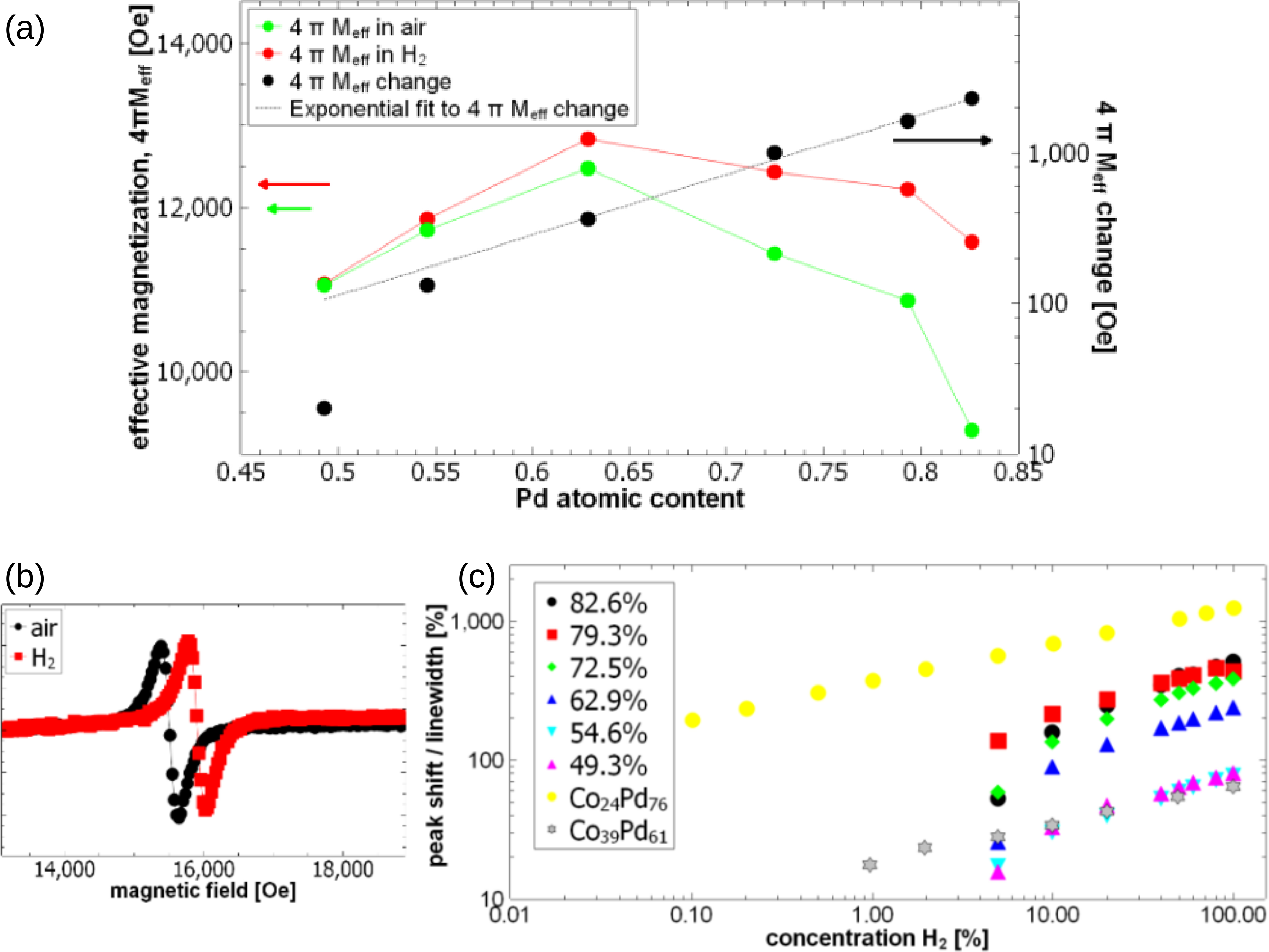}
\caption{\textbf{(a)}~The left axis shows the effective magnetisation $4\pi M_{eff}$, as defined in Sec.~\ref{sec:3.7}, in air (the green dots) and {\Ivan in an} H$_2$ atmosphere (the red dots) plotted as a function of the content of Pd in the material. The black dots on the right axis denote the difference in $4\pi M_{eff}$ plotted on a logarithmic scale (right axis), where the black dashed line is an exponential fit to the black dots. The red and green lines are the guides for the eye. \textbf{(b)}~Raw FMR traces at the microwave frequency 9\,GHz obtained for the sample with the composition Ni$_{26.1}$Co$_{11.0}$Pd$_{62.9}$ in H$_2$ environment (the red squares) and in air (the black dots). \textbf{(c)}~Normalised FMR peak shift for the Ni-Co-Pd films extracted from the raw FMR data at 9\,GHz as a function of the H$_2$ concentration. In the figure legend, the numbers from 49.3\% to 82.6\% denote the particular content of Pd in the Ni-Co-Pd films but the points labelled as "Co$_{24}$Pd$_{76}$" and "Co$_{39}$Pd$_{61}$" denote the data obtained for the respective reference Co-Pd alloy films. {\Ivan \copyright~2021 IEEE. Reprinted, with permission, from \cite{Sch21}.}.
\label{Fig14}}
\end{figure}

The effective magnetisation $4\pi M_{eff}$ and the gyromagnetic ratio $\gamma$ for the samples were extracted from the FMR measurements by fitting the obtained data with the corresponding Kittel equation \cite{Mak15}, where $4\pi M_{eff} = 4\pi M_s - H_{PMA}$ being $4\pi M_s$ the saturation magnetisation and $H_{PMA}$ the internal magnetic field due to PMA. It was found that the dependence of $4\pi M_{eff}$ on the Pd content is not linear but first it exhibits an increase and then a decrease [Fig.~\ref{Fig14}(a)]. The increase in $4\pi M_{eff}$ is consistent with that observed in the earlier studies of Ni-Fe-Pd alloys \cite{Ric79}, where this behaviour had been attributed to an increase in the concentration of FM species (Ni and Co) that effectively increases the saturation magnetisation of the material. The subsequent drop in $4\pi M_{eff}$ at around 63\% of Pd content in Fig.~\ref{Fig14}(a) was attributed to an effective decrease in the magnetic moment of Co caused by a lower magnetic moment of Ni. Furthermore, it is well-known that in close proximity of FM atoms, Pd atoms become magnetically polarised and that this contributes to the effective magnetisation $4\pi M_{eff}$. It also known that this effect is weaker in Ni-Pd systems than in Co-Pd ones \cite{Odo83}.

The measurements also revealed an {\Ivan exponential} increase in $4\pi M_{eff}$ for the Ni-Co-Pd alloys caused by hydrogenation, which is consistent with earlier similar observations for Pd/Co bilayers and Co-Pd alloy films. {\Ivan However, several other physical factors were identified to be responsible for this effect}. Firstly, absorption of H$_2$ modifies the electronic structure of the Co-Pd {\Ivan system}, thereby changing the magnetic environment of the atoms and eventually reducing the strength of the PMA \cite{Lue19, Mud17}. A plausible physical origin of this effect is a change in the Co-Pd orbital hybridisation \cite{Kim00} caused by hydrogen ions \cite{Lue17}. Since the addition of a small amount of Ni should not impact this physical mechanism noticeably, it was concluded that in the Ni-Co-Pd structures the change in $4\pi M_{eff}$ should also be due to a decrease in the strength of PMA. Secondly, the dilution of Pd with transition metals is known to reduce the ability of Pd to incorporate hydrogen into its crystal lattice \cite{Wan97}. In turn, a lower concentration of hydrogen in the crystal lattice has a smaller effect on the hybridisation of orbitals of Co and Pd atoms. Yet, a reduction of Pd content also implies that there would be fewer Co-Pd bonds that contribute significantly to PMA \cite{Kyu96}. Subsequently, it is legitimate to expect a smaller hydrogen-induced reduction in the strength of PMA at a smaller Pd content.

Figure~\ref{Fig14}(b) shows typical raw FMR traces measured in air and an H$_2$ environment, where a shift in the resonance field due to hydrogenation can be clearly seen. To demonstrate the dependence of the resonance shift on the concentration of H$_2$, in Fig.~\ref{Fig14}(c) this shift is normalised to the resonance peak linewidth and plotted as a function of the concentration of the gas. One can see that a higher Pd content in the alloy results in a more pronounced shift in the resonance field, even when the shift is normalised to the increased linewidth of the resonance peak. Only at the highest and lowest concentrations of H$_2$ considered in the experiment, the effect of a large linewidth compensates for a larger resonance peak shifts for the films with a high content of Pd. Interestingly, while the dependence of the normalised peak shift on the concentration of H$_2$ in Co-Pd is linear {\OKB on the log-log scale}, in Fig.~\ref{Fig14}(c) the data points for the Ni-Co-Pd films deviate downward with respect to a straight line.
 
Thus, this particular study have not produced a clear and unambiguous picture of the impact of Ni doping on the sensing properties of PdCo alloys. More investigations are needed to gain a deeper understanding of advantages and disadvantages of this more complex ternary-alloy system.  
  
\subsection{Advanced magneto-electronic sensors architectures: towards real-life applications\label{sec:3.8}}
As discussed in the previous sections, studies of both fundamental physical principles underlying the operation of magneto-electronic sensors and of their alternative active materials have been conducted with the ultimate goal of improving their sensitivity and chemical stability and thus getting them ready for real-life applications. In this section, we overview the results of several works that capitalise on the previous achievements and focus on further improvements of the sensor performance.
     
Firstly, it is worth noting the works where additional tests of fundamental physics of magneto-electronic sensors were conducted \cite{Kos17, Cau19}. For instance, in \cite{Cau19} an in operando depth-resolved study of the in-plane interfacial magnetisation of a Pd/Co film was conducted with a focus on PMA processes in the presence of H$_2$ gas. There, a neutron reflectometry technique was combined with the in situ FMR spectroscopy to explore the effect of the absorption of hydrogen at the Pd/Co interface on the {\OKB FMR}-condition during the cycling of H$_2$ gas. The experiment and modelling revealed that while the Pd layer expands when it is exposed to H$_2$, the in-plane magnetic moment of the Pd/Co film increases {\OKB slightly} and the interfacial PMA is reduced thereby affecting the FMR frequency. It was also noted that the use of non-standard experimental techniques also opens a potential alternative approach to magneto-electronic hydrogen sensing.

The effect of hydrogenation on the inverse spin Hall effect (iSHE) Pd/Co bilayer films was investigated in \cite{Wat18}, where the iSHE was driven by the FMR excited in the Co layer and measured as a dc voltage generated across the Pd layer. Recall that the generation of such a dc signal in the Pd layer provides a much safer approach to detection of H$_2$ gas than measuring the electric resistance of a Pd film. Indeed, the latter method requires application of a dc voltage to the active layer and, therefore, a fault in the electric circuit used to produce the applied voltage can potentially result in a spark leading to a very dangerous situation when H$_2$ is present. On the contrary, the iSHE voltage is fundamentally small and this naturally excludes any sparking. In \cite{Wat18}, it was demonstrated that in the presence of H$_2$ the iSHE peak shifts downward in applied field alongside the measured FMR absorption peak, which means that the sensitivity of the sensor exploiting the iSHE response should be the same as when the FMR response is used to measure the gas concentration. Moreover, an increase in the amplitude of the iSHE peak and its linewidth observed in the measurements could be used as an additional approach to hydrogen sensing. Note that a subsequent publication \cite{Wat20} on this effect, which predominantly focused on the physics underpinning the impact of hydrogen on the iSHE voltage, demonstrated that the incorporation of H atoms into the crystal lattice of Pd decreases the spin coherence length for spin-polarised electrons traversing the layer interface from Co to Pd, without affecting the spin-Hall angle for Pd.
 
\begin{figure}[t]
\includegraphics[width=12.5 cm]{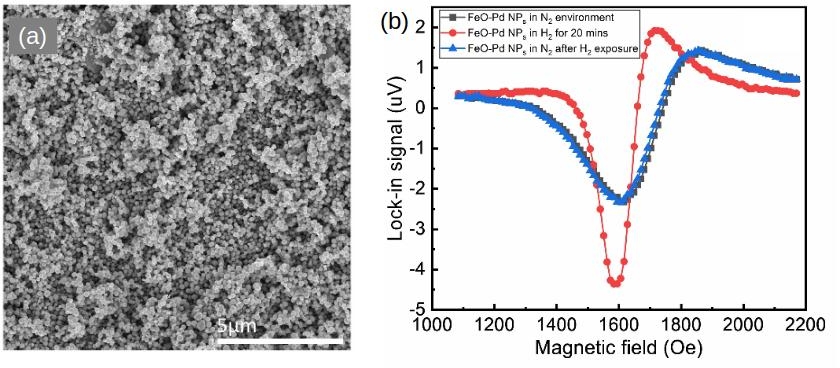}
\caption{\textbf{(a)}~Scanning electron microscopy image of the fabricated H$_2$-sensitive nanospheres and \textbf{(b)}~resonance peaks obtained as a result of in-plane FMR measurements in N$_2$ (squares), 3\% H$_2$ (circles) and post-H$_2$ exposure (triangles) atmospheres. Reprinted with permission from \cite{Sha_unpubl}.
\label{Fig15}}
\end{figure}

In the already mentioned work \cite{Sha_unpubl}, iron-oxide core/Pd shell nanospheres [Fig.~\ref{Fig15}(a)] were synthesised and the characterisation of their structural properties revealed significant changes in their magnetic properties caused by changes of the surrounding environment from pure N$_2$ gas to a mix of 3\% of H$_2$ in 97\% N$_2$. This property was used for efficient detection of H$_2$ gas at atmospheric pressure and room temperature. The nanospheres demonstrated superior FMR signal and reversible change in the magnetic properties upon hydrogenation compared with the thin-film and nanopatterned magneto-electronic sensors discussed in the previous sections. In particular, as one can see in Fig.~\ref{Fig15}(b), the nanospheres demonstrate a strong FMR signal with a small resonance linewidth, which is an important property for real-world applications of magneto-electronic hydrogen sensors. The resonance linewidth decreases in H$_2$ environment and the resonance position shifts to the lower static magnetic fields applied to the sensor. A strong increase in the FMR amplitude can also be seen. Significantly, a more than two-fold decrease in the linewidth is readily measurable using a relatively simple electronic circuit. This decrease is also accompanied by an increase in the amplitude of the FMR peak of about two times that is also easily measurable. It is noteworthy that these promising results were achieved at just 3\% of H$_2$ gas in the chamber used in the experiment, which was done for safety reason because we recall that the flammability threshold of hydrogen in air is 4\% and the respective explosivity threshold is 14\%. Thus, even better sensitivity figure-of-merit could be demonstrated at higher H$_2$ concentrations that may be encountered in a real-life situation, provided that all safety requirements are fulfilled in the experiment.

Finally in this section, we mention an on-chip integrated {\OKB FMR-based} hydrogen sensor employing Fe {\Ivan instead of Co} in the active layer \cite{Kha19}. Significantly, the linewidth of the FMR response of the Fe-based sensor is approximately 20\,Oe, which is much smaller than approximately 150\,Oe observed for a typical structure containing a Co layer. In the presence of H$_2$ gas, an FMR peak upshift of about 80\,Oe was observed. Although a similar behaviour has already been observed using Co-containing sensors, together with a narrow linewidth such a shift exhibited by the Fe-based sensor {\OKB may represent an} advantage for real-life applications since using it significantly simplifies the registration of the signal caused by the presence of H$_2$ gas.
   
\begin{figure}[t]
\includegraphics[width=7.5 cm]{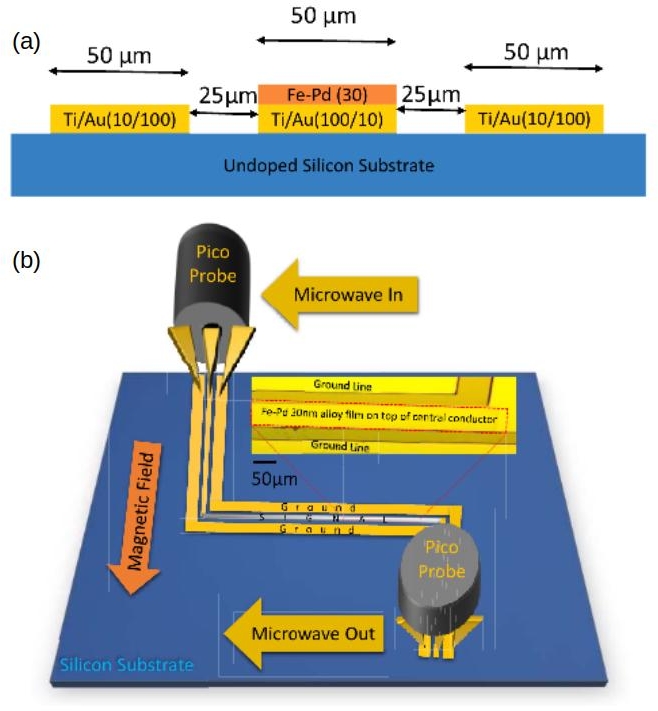}
\caption{\textbf{(a)}~Schematic of the cross-section of the fabricated microchip with a coplanar waveguide-like configuration. \textbf{(b)}~Schematic of a customised FMR characterisation station used to investigate microchips. The inset shows the magnified view of the central conductor of the coplanar waveguide. {\Ivan \copyright~2019 IEEE. Reprinted, with permission, from \cite{Kha19}.}
\label{Fig16}}
\end{figure}

The design of the sensor in \cite{Kha19} also addresses the established drawbacks of the previous generations of {\OKB FMR-based} sensors, including a relatively long response time, insufficient strength of the FMR signal and the use of sophisticated laboratory-level electronics to reliably detect the FMR response. In particular, to unleash the commercialisation potential of the sensor, miniaturised magneto-electronic sensors (microchips) were fabricated using a two-step lithography process, where a coplanar waveguide-like structure was created on top of the undoped silicon substrates using first the spin coating with negative photoresist and then patterning by mean of UV-photolithography. Then, layers of 10\,nm of Ti and 100\,nm of Au were deposited using an electron-beam assisted evaporative system and the resulting structures were transferred into an acetone bath to dissolve the photoresist layers and lift off the excess metal. In the second photolithography step, a 30-nm-thick Fe-Pd alloy film was deposited on top of the signal conductor of the coplanar line and the excess metal was lifted off in acetone bath. The resulting waveguide-like structure [Fig.~\ref{Fig16}(a)] was characterised in a custom-made pressure-tight chamber that houses the microscopic microwave probes Picoprobes\textregistered (GGB Industries). The probes were needed to connect the microchip in the laboratory setting [Fig.~\ref{Fig16}(b)]; we refer the interested reader to the review article \cite{Mak15}, where the advantages and physical peculiarities of the FMR spectroscopy using a coplanar waveguide-like structure are discussed in great detail. Overall, the fabricated microchip demonstrated a two-times stronger  FMR signal compared with that produced by a reference continuous film. Evidently, this is still work in the progress, but this very first result is already promising.

\section{Conclusions and outlook}
Thus, as discussed in this review article, there has been a significant progress in understanding the fundamental physical processes that underpin the operation of magneto-electronic hydrogen gas sensors. In particular, it has been demonstrated that the active materials used in these sensors exhibit high sensitivity to H$_2$ required for real-life applications, and that nanostructuring of these materials allows reducing the response time of the sensors and closely approaching it to the industry standard of one second. Hence, it is plausible to claim that magneto-electronic sensors have a strong potential to find their niche on the market of hydrogen gas sensors, which is estimated to reach a total market size of US\$773.724 million in 2024. Moreover, due to their specific properties and in particular intrinsically low fire risk, magneto-electronic sensors will be indispensable in applications where the safety is of especial importance, thereby becoming the technology of choice, for example, in mass transport hydrogen-powered vehicles. The capability of these systems to enable remote sensing \cite{Cha13} is also worth mentioning in this context.
 
However, extra research and development work is required for these sensors to reach the industry. Here, we outline some of the important milestones that will need to be completed to reach this goal.

Firstly, all these results have thus far been obtained for different sets of fabricated materials and none of the sample demonstrated a combination of all merits of the magneto-electronic sensor technology discussed in this review article. For example, while it has been shown that using alloyed active materials may be advantageous in many aspects, for instance, while measuring large H$_2$ concentrations, the actual list of potential candidate Pd-FM metal alloys is very long and the optimal alloy composition is yet to be established. In addition, more work has to be done to enable the control the FMR peak linewidth in the alloyed films. Furthermore, the effect of material nanostructuring on alloy-based structures has not been investigated yet. Consequently, a joint effort of physicists, material scientists, chemists and engineers is required to bridge this knowledge gap.
 
Thirdly, in the majority of research works completed so far, the static magnetisation vector of the investigated active structures naturally lies in their plane. Therefore, structuring such samples by forming long strips with a square nanoscale cross-section leads to a sensor operation in zero applied field, which is very beneficial for real-life applications since it simplifies the design of the sensor \cite{Lue16}. On the other hand, it was established that the sensitivity {\OKB of FMR-based sensors} to H$_2$ can be increased {\OKB manyfold} by {\OKB directing} the static magnetic moment of films perpendicular to the film plane \cite{Lue16_1}. However, to orient the magnetic moment in this direction, a strong magnetic bias field has to be applied to the material, thereby requiring large laboratory-level electromagnets that cannot be miniaturised to meet industry specifications. This challenge can be addressed using an approach where Co or Fe is diluted with Pd \cite{Lue19}, and therefore the static magnetic moment of the resulting alloy metal becomes so reduced that the static magnetic field needed to rotate the magnetic moment perpendicular to the plane decreases to a value that can be created using an inexpensive, compact and commercially available permanent magnet. Furthermore, based on purely theoretical considerations, it is plausible to assume that nanostructuring of alloyed materials should decrease the required bias field further, potentially to zero. It is also clear that in this case the geometry of nanowires will not be suitable since it intrinsically favours an in-plane orientation of the magnetic moment. Subsequently, other nanopatterning geometries promoting a perpendicular-to-plane orientation of the magnetic moment, and hence helping reduce the response of the sensor towards the industry standard, are required.
 
Another important problem that needs to be resolved is further simplification and cost reduction of the electronic equipment currently used to detect the FMR and iSHE responses of the active material of the sensor. Because the thickness of the films of the active material enabling optimal sensitivity can be just several nanometres, the volume of the material driven by the FMR is very small and therefore the rectified output voltages of the sensor are just of order of several micro-volts. One of the most feasible approaches to the resolution of this technological challenge is the use of very narrow microwave striplines as FMR transducers that can boost the signal produced by the sensor. As discussed in great detail in \cite{Mak15}, a large body of research has been devoted to the topic of stripline FMR spectroscopy in the non-sensor context. Therefore, applying the knowledge obtained in that field should help resolve problems in the area of {\OKB FMR based} sensors. It is also technically viable to integrate {\OKB FMR-based} sensors with inexpensive and readily available microwave oscillator circuits used in WiFi and mobile phone devices \cite{Lue16}. 

It is also worth mentioning other approaches to sensing the change in the magnetic properties of the films in the presence of hydrogen gas. The anomalous-Hall effect \cite{Ger17} is a way to replace the microwave probing signal with a dc one. However, this method requires injection of a dc current in the film that brings the same disadvantage of lower fire safety as the sensors based on measuring the dc resistance of Pd. Measuring the magneto-optical response (MOKE effect, Sec.~\ref{sec:MOKE}) of the magnetic materials brings the advantage of remote sensing of the gas. However, the rotation of light polarisation due to MOKE is usually small and therefore requires sensitive electronics, opto-electronics or magneto-plasmonics \cite{Mak15_review, Mak16_1} to detect the MOKE signal. Consequently, significant effort needs to be done to improve this alternative approach to magneto-electronic gas sensing, {\Ivan though some of these technologies have already been patented (see, e.g.,~\cite{LinPatent} and therefore they are expected to enter the service in commercial devices in the nearest future. One may also expect reports on more advanced magneto-electronic methods of hydrogen detection using, for example, novel phenomena and functionalities originating from the spin-orbit coupling (SOC). A fundamental building block of the SOC-based technology is the manipulation of spin-orbit torques (SOTs) that are often made of a heavy metal/FM heterostructures \cite{Hir15, Tak16, Hir20}. Therefore, recently it has been demonstrated that SOTs can also be used for hydrogen sensing \cite{An20, Hir21}. Therefore, the magneto-electronic sensors discussed in this article are likely to find important practical applications in the near future.} 

\vspace{6pt} 


\funding{\OKB We would like to thank the support by Australian Nuclear Science and Technology Organisation (ANSTO), grants No.~P4123, P4810 and P6126. We also acknowledge using the equipment of and receiving scientific and technical assistance from the Australian National Fabrication Facility and the Centre for Microscopy, Characterisation and Analysis of the University of Western Australia, a facility funded by the University, State and Commonwealth Governments. ISM has been supported by the Australian Research Council through the Future Fellowship (FT180100343) program.} 

\acknowledgments{We would like to thank the past and present students and postdocs at the University of Western Australia and ANSTO involved in the research work on hydrogen sensing as well as our colleagues for invaluable discussions. {\Ivan We also thank Dr Thomas Schefer for a careful reading of the manuscript and valuable suggestions.}}

\conflictsofinterest{The authors declare no conflict of interest.} 


\end{paracol}
\reftitle{References}


\externalbibliography{yes}
\bibliography{refs}

\end{document}